\documentclass[aps,prc,preprint,groupedaddress,nofootinbib]{revtex4-2}

\usepackage{graphicx}
\usepackage{amssymb}
\usepackage{amsmath}
\usepackage{color}
\usepackage{setspace}
\usepackage{amsfonts}

\makeatletter
	\def\erf{\mathop{\operator@font erf}\nolimits}
	\def\erfc{\mathop{\operator@font erfc}\nolimits}
	\def\Erf{\mathop{\operator@font Erf}\nolimits}
	\def\Shi{\mathop{\operator@font Shi}\nolimits}
	\def\Chi{\mathop{\operator@font Chi}\nolimits}
	\def\Ei{\mathop{\operator@font Ei}\nolimits}
	\def\cosec{\mathop{\operator@font cosec}\nolimits}
	\def\sech{\mathop{\operator@font sech}\nolimits}
	\def\cosech{\mathop{\operator@font cosech}\nolimits}
	\newcommand\hypgeo[2]{{}_{#1}{\operator@font F}_{#2}}
\makeatother

\begin{document}


\title{Angular correlations in neutron-gamma reactions}



\author{Vladimir Gudkov}
\email[]{gudkov@sc.edu}
\affiliation{Department of Physics and Astronomy, University of South Carolina, Columbia, South Carolina 29208, USA}

\author{Hirohiko M. Shimizu}
\email[]{shimizu@phi.phys.nagoya-u.ac.jp}
\affiliation{Department of Physics, Nagoya University, Nagoya 464-8602, Japan}


\date{\today}

\begin{abstract}
Angular correlations for low-energy neutron induced $\gamma$ reactions are calculated using nuclear reaction formalism for multi level neutron capture with arbitrary multiplicities of $\gamma$ transitions. We confirm previous calculations [V. V. Flambaum and O. P. Sushkov, Nucl. Phys.A
435
, 352 (1985)] for dipole transitions obtained with different formalism. The possible  applications of n-$\gamma$ correlations for studies of symmetry violations and electromagnetic transitions   are discussed.
\end{abstract}


\maketitle


\section{Introduction}

 Time reversal invariance violating (TRIV)
  effects in nuclear reactions which can be measured in the transmission of polarized neutrons through a polarized target\cite{Kabir:1982tp,Stodolsky:1982tp} attract interest since they   can be  enhanced \cite{Bunakov:1982is,Gudkov:1991qg} by a factor as large as $10^6$.
     Similar  enhancements \cite{Flambaum:1980hg,Sushkov:1982fa}  have been already observed for parity violating (PV) effects  in neutron transmission through nuclear targets. For example, the PV asymmetry in the $0.734 \ {\rm eV}$ $p$-wave resonance in $^{139}{\rm La}$ has been measured to be $(9.56 \pm 0.35)\cdot 10^{-2}$ (see, for example, Ref. \cite{Mitchell2001157} and references therein).
   The recent proposals for searches TRIV in neutron-nucleus scattering (see, for example, Ref. \cite{Bowman:2014fca} and references therein)  demonstrated the existence of a class of experiments that are free from false asymmetries, which have a discovery potential of  $10^2$ to $10^4$, being more sensitive to TRIV than current limits, because of the enhancement of TRIV effects  in the vicinity of neutron $p$-wave resonances in the compound nuclear system.
     Taking into account that different models of the CP violation may contribute differently to a particular T/CP-odd observable, which also may have  unknown theoretical uncertainties, TRIV nuclear effects could be considered   complementary   to  electric dipole moment measurements, whose status as null tests of T invariance is more widely known.

   It has been shown \cite{Gudkov:1990tb,Gudkov:1991qg} that the ratio of TRIV effects to PV effects  measured at the same $p$-wave resonance has  almost complete cancellation of nuclear reaction effects resulting in a ratio of TRIV to PV matrix elements taken between the same states with opposite parities. This eliminates a bulk of  nuclear   model uncertainties \footnote{We discuss here relations between experimental observables and nuclear matrix elements. For relations between nuclear matrix elements and nucleon coupling constants, which contain nuclear structure uncertainties, see for example \cite{Gudkov:1990tb,Gudkov:2013dp,Fadeev:2019bwc,Flambaum:2021zhn}.} in TRIV effects.
       The coefficient of proportionality  between the observables and the corresponding matrix elements involves neutron partial widths with different channel spins, which are unknown and must be determined from independent measurements of angular correlations in neutron induced reactions (see, for example \cite{Sushkov:1985ng}). Since these coefficients have a ``natural'' value of an order of 1, they were mostly ignored in previous studies of TRIV effects in neutron scattering. However,   it is very important to know the exact spin structure in the relation of PV and TRIV effects to be able to choose  the optimal  target  for TRIV experiments and for the further analysis of the experimental data.

 Due to complex structure of heavy nuclei it is impossible to calculate  neutron partial widths and their spin dependencies, however they can be measured from the set of experiments. One of the standard approach is extracting the values of neutron partial widths from the measurements of different angular correlations in neutron-induced $\gamma$ reactions. There are a number of recently performed \cite{Okudaira:2017dun,Yamamoto:2020xtz,Okudaira:2021svc,Koga:2022tb,Endo:2022nxr,Okudaira:2022wpa,Okuizumi:2024rkp} and ongoing experiments studying parameters of neutron resonances in neutron-$\gamma$ reactions.
  However, there is only one paper \cite{Sushkov:1985ng} with detailed and comprehensive calculations of these correlations. In general, since these parameters are very important, it is desirable to have other, preferably independent calculations to make assurance in the values of  neutron partial widths obtained from experiments. For the case of angular correlations of $\gamma$ reactions the situation is even more complicated, because of the existing number of confusions and inconsistencies in different calculations angular correlations. For example, for low energy n-$\gamma$ correlations it was claimed there was an observed discrepancy between calculations  \cite{Sushkov:1985ng} and measurements of the left-right asymmetry \cite{Alfim:1984} and   forward-backward asymmetry \cite{Alfim:1984} in radiative neutron capture at the vicinity of $^{117}$Sn 1.33 eV p-wave resonance (see, also \cite{Alfim:1991}).
 The analysis of this discrepancy was revisited by other authors \cite{Skoy:1991,Barabanov:1992}, but they also could not consistently describe all observable parameters, which raised the question about validity of  the process description. Therefore, we present correlations of angular calculations using the approach which is based on nuclear reaction theory and independent of the approach in \cite{Sushkov:1985ng}.

\section{Angular correlations for spin-1 massive particles in neutron-induced reactions}

First we derive angular correlations  for spin-1 ($j=1$) massive particles in neutron (spin $s =1/2$) induced  reactions on a target with spin $I$,  and with  the residual nucleus with spin $I_F$.
 Choosing quantization axis $z$ along the direction of neutron momentum,   one can write a  general expression for matrix elements of the  reaction matrix $\hat{\mathbb{T}}$ \cite{Baldin:1961}  as
\begin{eqnarray}\label{genMat}
 \nonumber
  2\pi i
  \left\langle \mu_j M_F \middle\vert \mathbb{T} \middle\vert \mu M_I \right\rangle
  &=& \sum_{JM_Jlml^{\prime} m^{\prime} S m_s S^{\prime} m^{\prime}_s}Y_{l^{\prime} m^{\prime}}(\theta ,\phi)
  	\left\langle j \mu_j I_F M_F \middle\vert S^{\prime} m^{\prime}_s \right\rangle
	\left\langle   l^{\prime} m^{\prime} S^{\prime} m^{\prime}_s \middle\vert JM \right\rangle\\
  &\times&
  	\left\langle S^{\prime} l^{\prime} \alpha^{\prime} \middle\vert R^J \middle\vert S l \alpha\right\rangle
	\left\langle JM \middle\vert   l m S m_s \right\rangle
	\left\langle S m_s \middle\vert s \mu I M_I \right\rangle
	Y^*_{lm} (0 ,0) ,
 \end{eqnarray}
where primed parameters correspond to outgoing channel, and angles $(\theta ,\phi)$ describe the direction of the outgoing particle $j$.
 The matrix $\hat{R}$ is related to the scattering matrix  $\hat{\mathbb{S}}$ as $\hat{R}=\hat{1}-\hat{\mathbb{S}}$ which  in the integral of motion representation \cite{Baldin:1961} is
\begin{equation}\label{Smat}
\left\langle S^{\prime} l^{\prime} \alpha^{\prime} \middle\vert \mathbb{S}^J \middle\vert S l \alpha \right\rangle
\delta_{JJ^{\prime}}\delta_{MM^{\prime}}\delta(E^{\prime} -E) ,
\end{equation}
where $J$ and $M$ are the total spin and its projection, $S$ is the channel spin, $l$ is the orbital momentum, and $\alpha$ represents the other internal quantum numbers.
Let us consider the case with unpolarized target and residual nucleus.
In that case, tensor polarizations  of  rank $q$ of the target $\tau^I_{q \kappa}$ and of the final nucleus $\tau^{I_F}_{q \kappa}$    vanish except for $q=\kappa =0$.
 Then we need to describe only neutron
 polarization, whose     density matrix in terms of the statistical tensors $\tau_{q \kappa}$ can be written as
 \begin{equation}\label{denMart}
   \rho_{\mu\mu^\prime}=\sum_{q \kappa}\sqrt{2q+1}\left\langle I \mu q \kappa \middle\vert I \mu^\prime \right\rangle \tau_{q \kappa}.
 \end{equation}
Therefore, the expression for corresponding statistical tensors for the outgoing particle are
 \begin{equation}\label{Tens}
   \left\langle T_{QK} \right\rangle=Tr(\hat{\mathbb{T}}\rho \hat{\mathbb{T}}^{\dag}),
 \end{equation}
  which can be written explicitly as
   \begin{eqnarray}\label{TauClebsch}
     \left\langle  T_{QK} \right\rangle
  &=&\frac{\sqrt{(2Q+1)}}{4\pi}\sum_{JMll^{\prime} m^{\prime} S m_s S^{\prime} m^{\prime}_s \mu q\kappa L M_L}\tau_{q \kappa} Y_{L M_L}(\theta ,\phi) \\
   \nonumber
   &\times&
  \sqrt{\frac{(2l_1+1)(2l_2+1)(2l^{\prime}_1+1)(2l^{\prime}_2+1)(2q+1)}{4\pi (2L+1)}} (-1)^{m^{\prime}_1}
   \\
   \nonumber
   &\times&
   \left\langle L0 \middle\vert l^{\prime}_1 0 l^{\prime}_2 0 \right\rangle
  \left\langle LM_L \middle\vert l^{\prime}_1 -m^{\prime}_1 l^{\prime}_2 m^{\prime}_2 \right\rangle
    \left\langle j \mu_{j2} QK\vert j \mu_{j1} \right\rangle
  	\left\langle j \mu_{j1} I_F M_F \middle\vert S^{\prime}_1 m^{\prime}_{s1} \right\rangle
  \left\langle j \mu_{j2} I_F M_F \middle\vert S^{\prime}_2 m^{\prime}_{s2} \right\rangle \\
   \nonumber
      &\times&
  	\left\langle   l^{\prime}_1 m^{\prime}_1 S^{\prime}_1 m^{\prime}_{s1} \middle\vert J_1M_1 \right\rangle
  \left\langle   l^{\prime}_2 m^{\prime}_2 S^{\prime}_2 m^{\prime}_{s2} \middle\vert J_2M_2 \right\rangle
  \left\langle S^{\prime}_1 l^{\prime}_1 \alpha^{\prime} \middle\vert R^{J_1} \middle\vert S_1 l_1 \alpha\right\rangle^*
   \left\langle S^{\prime}_2 l^{\prime}_2 \alpha^{\prime} \middle\vert R^{J_2} \middle\vert S_2 l_2 \alpha\right\rangle
   \\
   \nonumber
  &\times&
  		\left\langle J_1M_1 \middle\vert   l_1 0 S_1 m_{s1} \right\rangle
  \left\langle J_2M_2 \middle\vert  l_2 0 S_2 m_{s2}  \right\rangle
  	\left\langle S_1 m_{s1} \middle\vert s \mu_1 I M_I \right\rangle
  \left\langle S_2 m_{s2} \middle\vert s \mu_2 I M_I \right\rangle
  \left\langle s \mu_{2} q \kappa\vert s \mu_{1} \right\rangle
	.
 \end{eqnarray}

To sum this expression over magnetic numbers it is convenient to start from three last Clebsch-Gordan coefficients. Then, using the identity (19) of \cite{RevModPhys.24.249} one can transform their sum to
\begin{equation}\label{sum12}
 (-1)^{1/2-I+\kappa-M_1}\sqrt{\frac{2(2S_1+1)(2S_2+1)}{(2q+1)}} \left\langle q -\kappa \middle\vert S_1 -M_1 S_2 M_2 \right\rangle
 W(S_1\frac{1}{2} S_2 \frac{1}{2};Iq),
\end{equation}
where $ W(S_1\frac{1}{2} S_2 \frac{1}{2};Iq)$ is Racah coefficient. Using the same identity one can transform the sum of three last Clebsch-Gordan coefficients in the third line of eq.(\ref{TauClebsch}) to
\begin{equation}\label{sum1}
  (-1)^{I_F +2S^{\prime}_2-2S^{\prime}_1-K-m^{\prime}_{s1}+1 }\sqrt{\frac{3(2S^{\prime}_2+1)(2S^{\prime}_1+1)}{(2Q+1)} }
  \left\langle Q -K \middle\vert S^{\prime}_1 - m{\prime}_{s1} S^{\prime}_2  m{\prime}_{s2}  \right\rangle
  W(S^{\prime}_1 1S^{\prime}_2 1;I_F Q),
\end{equation}
where the value $j=1$ used explicitly. Now, substituting these sums in eq.(\ref{TauClebsch}) we obtain
\begin{eqnarray}\label{TauClebsch2}
     \left\langle  T_{QK} \right\rangle
  &=&\frac{1}{4\pi}\sum_{JMll^{\prime} m^{\prime} S m_s S^{\prime} m^{\prime}_s q\kappa L M_L}\tau_{q \kappa} Y_{L M_L}(\theta ,\phi) (-1)^{-I+I_F-1/2+S_1+ 2S^{\prime}_2 -2S^{\prime}_1+\kappa -K}\\
   \nonumber
   &\times&
  \sqrt{\frac{3(2l_1+1)(2l_2+1)(2l^{\prime}_1+1)(2l^{\prime}_2+1) (2S_1+1)(2S_2+1)(2S^{\prime}_1+1)(2S^{\prime}_2+1)}{2\pi (2L+1)}}
   \\
   \nonumber
   &\times&
      \left\langle S^{\prime}_1 l^{\prime}_1 \alpha^{\prime} \middle\vert R^{J_1} \middle\vert S_1 l_1 \alpha\right\rangle^*
   \left\langle S^{\prime}_2 l^{\prime}_2 \alpha^{\prime} \middle\vert R^{J_2} \middle\vert S_2 l_2 \alpha\right\rangle \\
   \nonumber
   &\times&
   \left\langle L0 \middle\vert l^{\prime}_1 0 l^{\prime}_2 0 \right\rangle
  W(S_1\frac{1}{2} S_2 \frac{1}{2};Iq)W(S^{\prime}_1 1S^{\prime}_2 1;I_F Q)
    \\
   \nonumber
      &\times& (-1)^{m^{\prime}_1-m^{\prime}_{s1}-M_1}
       \left\langle LM_L \middle\vert l^{\prime}_1 -m^{\prime}_1 l^{\prime}_2 m^{\prime}_2 \right\rangle
  	\left\langle   l^{\prime}_1 m^{\prime}_1 S^{\prime}_1 m^{\prime}_{s1} \middle\vert J_1M_1 \right\rangle
  \left\langle   l^{\prime}_2 m^{\prime}_2  S^{\prime}_2 m^{\prime}_{s2} \middle\vert J_2M_2 \right\rangle
     \\
   \nonumber
  &\times&
  		\left\langle J_1M_1 \middle\vert   l_1 0 S_1 m_{s1} \right\rangle
  \left\langle J_2M_2 \middle\vert   l_2 0 S_2 m_{s2} \right\rangle
  \left\langle q -\kappa \middle\vert S_1 -M_1 S_2  M_2 \right\rangle	
 \left\langle Q -K \middle\vert S^{\prime}_1 -m{\prime}_{s1} S^{\prime}_2 m{\prime}_{s2}  \right\rangle
	.
 \end{eqnarray}

The sum over  magnetic numbers in two last lines in eq.(\ref{TauClebsch2}) can be evaluated using result (3.1) of \cite{Simon:gamma} as
\begin{eqnarray} \label{sum7}
 \nonumber
   & & \sum_{r,R} (-1)^{S^{\prime}_2+S_2+l_2+l^{\prime}_2}
       (2J_1+1)(2J_2+1) \sqrt{(2Q+1)(2q+1)(2L+1)(2r+1)} \\
   &\times& \left\langle r0 \middle\vert l_1 0  l_2 0  \right\rangle
   \left\langle R \kappa \middle\vert r 0  q \kappa  \right\rangle
   \left\langle R \rho \middle\vert L (\rho-K)  QK  \right\rangle
   \begin{Bmatrix}
  J_1 & l_1 & S_1 \\
   R & r & q \\
   J_2 & l_2 & S_2
  \end{Bmatrix}
   \begin{Bmatrix}
  J_1 & l^{\prime}_1 & S^{\prime}_1 \\
   R & L & Q \\
   J_2 & l^{\prime}_2 & S^{\prime}_2
  \end{Bmatrix} ,
     \end{eqnarray}
where two last terms are $9j$ symbols.

Substituting this expression in eq.(\ref{TauClebsch2}) we obtain a general expression for statistical tensors  for the case of massive particle with spin 1:
\begin{eqnarray}\label{TauPart}
     \left\langle  T_{QK} \right\rangle
  &=&\frac{\sqrt{(2Q+1)}}{4\pi\sqrt{2\pi}}\sum_{Jll^{\prime}  S  S^{\prime} Rr q \kappa  L \rho}\tau_{q \kappa} Y_{L K-\kappa}(\theta ,\phi) \\
  \nonumber
  &\times&
  (-1)^{-I+I_F-1/2+S_2+ 3S^{\prime}_2 -2S^{\prime}_1+l^{\prime}_2+l_2+\kappa -K}
  (2J_1+1)(2J_2+1) \\
   \nonumber
   &\times&
  \sqrt{3(2l_1+1)(2l_2+1)(2l^{\prime}_1+1)(2l^{\prime}_2+1)(2q+1) (2r+1) (2S_1+1)(2S_2+1)(2S^{\prime}_1+1)(2S^{\prime}_2+1)}
   \\
   \nonumber
   &\times&
      \left\langle S^{\prime}_1 l^{\prime}_1 \alpha^{\prime} \middle\vert R^{J_1} \middle\vert S_1 l_1 \alpha\right\rangle^*
   \left\langle S^{\prime}_2 l^{\prime}_2 \alpha^{\prime} \middle\vert R^{J_2} \middle\vert S_2 l_2 \alpha\right\rangle \\
   \nonumber
   &\times&
   \left\langle L0 \middle\vert l^{\prime}_1 0 l^{\prime}_2 0 \right\rangle
  W(S_1\frac{1}{2} S_2 \frac{1}{2};Iq)W(S^{\prime}_1 1S^{\prime}_2 1;I_F Q)
    \\
   \nonumber
   &\times&
     \left\langle r0 \middle\vert l_1 0  l_2 0  \right\rangle
   \left\langle R \kappa \middle\vert r 0  q \kappa  \right\rangle
   \left\langle R \rho \middle\vert L (\rho-K)  QK  \right\rangle
   \begin{Bmatrix}
  J_1 & l_1 & S_1 \\
   R & r & q \\
   J_2 & l_2 & S_2
  \end{Bmatrix}
   \begin{Bmatrix}
  J_1 & l^{\prime}_1 & S^{\prime}_1 \\
   R & L & Q \\
   J_2 & l^{\prime}_2 & S^{\prime}_2
  \end{Bmatrix}
	.
 \end{eqnarray}

\section{Angular correlations for neutron-$\gamma$ reactions}
To transform expressions involving spin-1 massive particles to the case of $\gamma$ rays we need to change the representation for the final state of $\hat{R}$ matrix from the channel spin to the multipole representation for $\gamma$-rays (see, for example \cite{Simon:gamma,Baldin:1961} and references therein)
\begin{equation}\label{transform}
 \left\langle S^{\prime} l^{\prime} \alpha^{\prime} \middle\vert R^{J} \middle\vert S l \alpha\right\rangle \rightarrow \sum_{gp}\left\langle pg \alpha^{\prime} \middle\vert R^{J} \middle\vert S l \alpha\right\rangle (gp \vert S^{\prime} l^{\prime}),
\end{equation}
where $g$ is the total angular momentum of $\gamma$ related to the electromagnetic multipole transition  with the "parity" $p$: $p=0$ for magnetic radiation and $p=1$ for electric radiation.

It should be noted that application of this transformation was a source of multiple confusions and/or mistakes in  calculations of angular correlations in nuclear reactions with photons (see, for example \cite{PhysRevC.20.453,PhysRevC.20.389,RevModPhys.39.306,RevModPhys.30.257,Simon:gamma,Baldin:1961} and references therein). The source of the confusions is related to the definition of the transformation function $(gp \vert S^{\prime} l^{\prime})$, which transforms wave function of massive particle with spin 1 to photon wave function. The photon wave functions must satisfy the conditions of the transversality and orthogonality, have defined parity and correctly defined phase related to time reversal transformation \cite{Baldin:1961}. This transformation function can be written as \cite{Baldin:1961}
\begin{equation}\label{gp}
(gp \vert S^{\prime} l^{\prime})=-\sqrt{2}(-1)^p\sqrt{(2g+1)(2S^{\prime}+1)}
W(l^{\prime}1JI_F;gS^{\prime})
\left\langle l^{\prime} 0 \middle\vert g -1 1 1   \right\rangle
\delta (l^{\prime},p),
\end{equation}
where, for $p=0$
\begin{equation}\label{delta0}
 \delta (l^{\prime},p)=\begin{cases}
    1, & \text{if $l^{\prime}=g$}\\
    0, & \text{if $l^{\prime}\neq g$},
  \end{cases}
\end{equation}
and for $p=1$
\begin{equation}\label{delta1}
 \delta (l^{\prime},p)=\begin{cases}
    0, & \text{if $l^{\prime}=g$}\\
    1, & \text{if $l^{\prime}\neq g$}.
  \end{cases}
\end{equation}

Unfortunately, in \cite{Baldin:1961} and many other publications authors did not pay attention to the uniqueness \cite{PhysRevC.20.453,PhysRevC.20.389} of the definition of this transformation function, which can be also defined by eq.(\ref{gp}) without  the factor $(-1)^p$. Both of these definitions (with and without the factor $(-1)^p$) are valid and may lead to correct results provided they applied consistently for calculations of all parameter in the description n-$\gamma$ reactions. However, the explicit expressions of differential cross sections for these two choices  may look different  because some angular correlations may have different signs. To show how the choice of  different transformation functions affect differential cross sections we follow results of  \cite{PhysRevC.20.453,PhysRevC.20.389,RevModPhys.39.306} (for more details, see also references in \cite{PhysRevC.20.453,PhysRevC.20.389,RevModPhys.39.306}). First, we recall that  general solutions of the wave equation  for transverse components of electromagnetic vector potential for electric $\vec{A}_{el}$ and magnetic $\vec{A}_{mag}$ multipoles, which corresponds to photon wave functions, are related by the following curl transformation
\begin{equation}\label{curl}
\nabla \times \vec{A}_{el}=\pm \vec{A}_{mag}.
\end{equation}
The sign on the right side of the equation cannot be fixed by applying the conditions discussed above, but  often a  positive sign is chosen. However, since the choice of this sign results in different relative  signs of electric and magnetic photon wave functions, the transformation functions must have different signs for these two choices. In particular, eq.(\ref{gp}) given in \cite{Baldin:1961} corresponds to minus sign in eq.(\ref{curl}), but by dropping the factor $(-1)^p$ in eq.(\ref{gp}) one get the transformation function which corresponds  to  positive sign in eq.(\ref{curl}).

 Now we show, that both choices (provided we know the choice of the sign and treat it consistently) will give the same result for nuclear reactions with photons. One can see that   the product of the $\hat{R}$ matrix and the transformation function in eq.(\ref{transform}) effectively results in a product of the amplitude of $\gamma$ transition $\Gamma^{1/2}_{\gamma} \sim <\psi_{f}| \hat{A}_{el,mag}|\psi_{in} > $ between initial $\psi_{in}$ and final $\psi_{f}$ states and the transformation function
\begin{equation}\label{transgamma}
\left\langle pg \alpha^{\prime} \middle\vert R^{J} \middle\vert S l \alpha\right\rangle (gp \vert S^{\prime} l^{\prime})  \rightarrow  <\psi_{f}| \hat{A}_{el,mag}|\psi_{in} >   (gp \vert S^{\prime} l^{\prime}).
\end{equation}
Thus, the change  of the sign of the electromagnetic operator compensate the change of the sign in the transformation function. This does not lead to a confusion where, like in atomic physics, for example, we can calculate amplitudes of electromagnetic transitions explicitly.

The situation is different for  most cases in nuclear physics where $\gamma$ decay amplitude cannot be calculated, but rather considered   as a parameter, which can be extracted from the experimental results. In that case,  the right side of eq.(\ref{transgamma})   written explicitly in terms of amplitudes of $\gamma$ widths  $\Gamma^{1/2}_{\gamma}$
\begin{equation}\label{transnucl}
  <\psi_{f}| \hat{A}_{el,mag}|\psi_{in} >   (gp \vert S^{\prime} l^{\prime}) \rightarrow \Gamma^{1/2}_{\gamma}(gp \vert S^{\prime} l^{\prime}) ,
\end{equation}
 does not  ``transparently absorb'' the choosen sign of the transformation function. Therefore, a different choice of the sign of the transformation function   leads to different signs of angular correlations in differential cross sections written  in terms of $\Gamma^{1/2}_{\gamma}$'s for the terms which contain   interference of electric and magnetic transitions. The terms without such interferences have the same signs irrespective of the choice of the transformation function.  This fact leads to the confusions  in the literature  in description of angular correlations in n-$\gamma$ reactions, which  could  easily be resolved if the discussed sign choices were explicitly mentioned  in the publications. Unfortunately, this is not the case.

As an important example for study of spectroscopic parameters of low energy neutron resonances, one can mention the calculations of angular correlations in n-$\gamma$ reactions in \cite{Sushkov:1985ng}, which have been done in different formalism without the discussed transformation functions,   and the  calculations using transformation functions \cite{Baldin:1961} in nuclear reaction formalism.  Analysing  the results of these calculations, one can see that  the first calculations have been done with the choice of a plus sign in eq.(\ref{curl}), which  usually used in atomic physics. The second approach corresponds to the choice of minus sign. Since the  choices of the signs have not been explicitly mentioned in both of these publications, the formal comparison leads to the illusion that at least one of the calculations is not correct.

We are working in nuclear reaction formalism (see, for example \cite{Baldin:1961}). However, to be able to compare our results directly with the first (and only, to our knowledge) detailed and comprehensive calculations \cite{Sushkov:1985ng} of n-$\gamma$ angular correlations in low energy nuclear reactions, we choose  the transformation function defined by eq.(\ref{gp}), but without  the factor $(-1)^p$. As we can see from the above discussion, the other choice will results in the additional phase $(-1)^{(p_1+p_2)}$ in following expressions providing an opposite sign for the terms with interference of electric and magnetic transitions.

For further calculations it is convenient to use the identity \cite{PhysRev.95.440}
\begin{equation}\label{pariry}
\left\langle l^{\prime} 0 \middle\vert g -1 1 1   \right\rangle
\delta (l^{\prime},p) =\frac{1}{2} \left[\left\langle l^{\prime} 0 \middle\vert g -1 1 1   \right\rangle - (-1)^p \left\langle l^{\prime} 0 \middle\vert g 1 1 -1   \right\rangle       \right].
\end{equation}
Substituting (\ref{transform}) in eq.(\ref{TauPart}) we obtain the $\left\langle  T_{QK} \right\rangle$ for n-$\gamma$ reactions.
Since matrix elements $\left\langle pg \alpha^{\prime} \vert R^{J} \vert S l \alpha \right\rangle $  do not depend on channel spins $S^{\prime}_1$ and $S^{\prime}_2$, we can sum over them to simplify the expression. To do this, first rewrite last coefficient in eq.(\ref{TauPart}) as \cite{Simon:part}
\begin{equation}\label{9jtoW}
   \begin{Bmatrix}
  J_1 & l^{\prime}_1 & S^{\prime}_1 \\
   R & L & Q \\
   J_2 & l^{\prime}_2 & S^{\prime}_2
  \end{Bmatrix} = (-1)^{\Sigma}
  \sum_x (2x+1)W( l^{\prime}_1 R S^{\prime}_1 J_2;x J_1)W(R l^{\prime}_1 Q l^{\prime}_2; x L )W(J_2 S^{\prime}_1 l^{\prime}_2 Q ;x S^{\prime}_2),
\end{equation}
where $\Sigma = J_1+J_2+S^{\prime}_1+S^{\prime}_2+l^{\prime}_1+l^{\prime}_2+R+L+Q $.

Now we can calculate the sum  over $S^{\prime}_2$, which includes a product of three Racah coefficients: the last one from from eq.(\ref{9jtoW}), another one  from the expression in eq.(\ref{gp}), and the last coefficient from the 5th line of eq.(\ref{TauPart}), multiplied by the term $\sqrt{(2S^{\prime}_2+1)}$ from 3rd line of eq.(\ref{TauPart}). Using  identity (17) of \cite{RevModPhys.24.249} this sum can be written as
\begin{equation}\label{s2sum}
  (-1)^{2S^{\prime}_1+1-Q-x-g_1}W(S^{\prime}_1 1 J_2 g_2;I_F x)W(Q 1 l^{\prime}_2 g_2; 1 x).
\end{equation}

In a similar way, by collecting three Racah coefficients which depends on $S^{\prime}_1$ from the above expression and eqs.(\ref{gp}) and (\ref{9jtoW}) with additional terms from eq.(\ref{TauPart}), and by applying identity (17) of \cite{RevModPhys.24.249}, we can write the sum over $S^{\prime}_1$ as
\begin{equation}\label{s1sum}
  (-1)^{J_1+J_2 -I_F -x -g_1}W(J_1 R I_F g_2;J_2 g_1)W(l^{\prime}_1 R 1 g_2; x g_1).
\end{equation}

Then, substituting eq.(\ref{transform}) into eq.(\ref{TauPart}) with results on the summations of (\ref{s2sum}) and (\ref{s1sum}), and using (\ref{9jtoW}), we can write an expression for n-$\gamma$ reactions as
\begin{eqnarray}\label{TauNgammalpr}
     \left\langle  T_{QK} \right\rangle
  &=&\frac{\sqrt{3(2Q+1)}}{2\pi\sqrt{2\pi}}\sum_{Jl l^{\prime} S  g p  Rr q \kappa  L \rho} Y_{L K-\kappa}(\theta ,\phi) \sqrt{(2q+1)  (2r+1)}\\
  \nonumber
  &\times&
  (-1)^{2(J_1+J_2)+S_2-I+1/2+l^{\prime}_2+ l_2-g_1 -g_2 -Q+\kappa -K}
  (2J_1+1)(2J_2+1) \\
   \nonumber
   &\times&
  \sqrt{(2l_1+1)(2l_2+1)(2l^{\prime}_1+1)(2l^{\prime}_2+1) (2S_1+1)(2S_2+1)(2g_1+1)(2g_2+1)}
   \\
   \nonumber
   &\times&
      \left\langle p_1 g_1 \alpha^{\prime} \middle\vert R^{J_1} \middle\vert S_1 l_1 \alpha\right\rangle^*
   \left\langle p_2 g_2 \alpha^{\prime} \middle\vert R^{J_2} \middle\vert S_2 l_2 \alpha\right\rangle
   \left\langle l^{\prime}_1 0 \middle\vert g_1 -1 1 1   \right\rangle
\delta (l^{\prime}_1,p_1)
\left\langle l^{\prime}_2 0 \middle\vert g_2 -1 1 1   \right\rangle
\delta (l^{\prime}_2,p_2)
    \\
   \nonumber
   &\times&
     W(S_1\frac{1}{2} S_2 \frac{1}{2};Iq)
 W(J_1 R I_F g_2;J_2 g_1)
 \left\langle L0 \middle\vert l^{\prime}_1 0 l^{\prime}_2 0 \right\rangle
     \left\langle r0 \middle\vert l_1 0  l_2 0  \right\rangle
      \\
   \nonumber
   &\times&
      \left\langle R \kappa \middle\vert r 0  q \kappa  \right\rangle
   \left\langle R \rho \middle\vert L (\rho-K)  QK  \right\rangle
   \begin{Bmatrix}
  J_1 & l_1 & S_1 \\
   R & r & q \\
   J_2 & l_2 & S_2
  \end{Bmatrix}
    \begin{Bmatrix}
  g_1 & l^{\prime}_1 & 1 \\
   R & L & Q \\
   g_2 & l^{\prime}_2 & 1
  \end{Bmatrix}
   	.
 \end{eqnarray}

Now we can sum over $l^{\prime}_1$ and $l^{\prime}_2$ in the above expression using eq.(\ref{pariry}) and identity (41) in Sec. 8.7.6 of \cite{Varshalovich}. Then, defining the function
\begin{eqnarray}\label{EmQ}
  Em&(&g_1,p_1,g_2,p_2,L,R,Q) = \frac{(-1)^{R+g_1-g_2}}{4\sqrt{(2R+1)(2Q+1)}} \\
  \nonumber
  &\times& \left(
  \left\langle L0 \middle\vert R -2 Q 2 \right\rangle
  \left\langle R 2 \middle\vert g_1 1 g_2 1 \right\rangle
  \left\langle Q 2 \middle\vert 1 1 1 1 \right\rangle (-1)^Q [(-1)^{p_2+g_2} + (-1)^{p_1-g_1-L}]
   \right.
   \\
  \nonumber
  &-& \left.
    \left\langle L0 \middle\vert R 0 Q 0 \right\rangle
  \left\langle R 0 \middle\vert g_1 -1 g_2 1 \right\rangle
  \left\langle Q 0 \middle\vert 1 -1 1 1 \right\rangle (-1)^{-L} [(-1)^{-g_1} + (-1)^{p_1+p_2+g_2 +Q-R}]
  \right),
\end{eqnarray}
we can write the final expression for n-$\gamma$ reactions as
\begin{eqnarray}\label{TauNgamma}
     \left\langle  T_{QK} \right\rangle
  &=&\frac{\sqrt{3(2Q+1)}}{2\pi\sqrt{2\pi}}\sum_{Jl  S  g p  Rr q \kappa  L \rho}\tau_{q \kappa} Y_{L K-\kappa}(\theta ,\phi) \sqrt{(2q+1)  (2r+1)}\\
  \nonumber
  &\times&
  (-1)^{2(J_1+J_2)+S_2-I+1/2+ l_2-g_1 -g_2 -Q+\kappa -K}
  (2J_1+1)(2J_2+1) \\
   \nonumber
   &\times&
  \sqrt{(2l_1+1)(2l_2+1) (2S_1+1)(2S_2+1)(2g_1+1)(2g_2+1)}
   \\
   \nonumber
   &\times&
      \left\langle p_1 g_1 \alpha^{\prime} \middle\vert R^{J_1} \middle\vert S_1 l_1 \alpha\right\rangle^*
   \left\langle p_2 g_2 \alpha^{\prime} \middle\vert R^{J_2} \middle\vert S_2 l_2 \alpha\right\rangle
   Em(g_1,p_1,g_2,p_2,L,R,Q)
    \\
   \nonumber
   &\times&
     W(S_1\frac{1}{2} S_2 \frac{1}{2};Iq)
 W(J_1 R I_F g_2;J_2 g_1)
     \left\langle r0 \middle\vert l_1 0  l_2 0  \right\rangle
      \\
   \nonumber
   &\times&
      \left\langle R \kappa \middle\vert r 0  q \kappa  \right\rangle
   \left\langle R \rho \middle\vert L (\rho-K)  QK  \right\rangle
   \begin{Bmatrix}
  J_1 & l_1 & S_1 \\
   R & r & q \\
   J_2 & l_2 & S_2
  \end{Bmatrix}
   	.
 \end{eqnarray}

Let us consider  the case without polarization of $\gamma$-rays. In this case $Q=K=0$, which leads to the following simplifications of the  Em function
\begin{eqnarray}\label{Em}
  Em&(&g_1,p_1,g_2,p_2,L,R,0) = \frac{(-1)^{1-g_2} \delta_{LR}}{4\sqrt{3(2L+1)}} \\
  \nonumber
   &\times& \left(
      \left\langle L 0 \middle\vert g_1 -1 g_2 1 \right\rangle
  [1 + (-1)^{p_1+p_2+g_1+g_2-L}]
  \right),
\end{eqnarray}
 and, as a consequence,  the expression for $\left\langle  T_{00} \right\rangle$ can be written as
 \begin{eqnarray}\label{TauNgamma00}
     \left\langle  T_{00} \right\rangle
  &=&\frac{1}{8\pi\sqrt{2\pi}}\sum_{Jl  S  g p  r q \kappa  L }\tau_{q \kappa} Y_{L -\kappa}(\theta ,\phi) \sqrt{\frac{(2q+1)  (2r+1)}{(2L+1)}}\\
  \nonumber
  &\times&
  (-1)^{2(J_1+J_2)+S_2-I+3/2+ l_2-l_1-g_1 +\kappa}
  (2J_1+1)(2J_2+1) \\
   \nonumber
   &\times&
  \sqrt{(2l_1+1)(2l_2+1) (2S_1+1)(2S_2+1)(2g_1+1)(2g_2+1)}
   \\
   \nonumber
   &\times&
      \left\langle p_1 g_1 \alpha^{\prime} \middle\vert R^{J_1} \middle\vert S_1 l_1 \alpha\right\rangle^*
   \left\langle p_2 g_2 \alpha^{\prime} \middle\vert R^{J_2} \middle\vert S_2 l_2 \alpha\right\rangle
       \\
   \nonumber
   &\times&
    \left\langle r0 \middle\vert l_1 0  l_2 0  \right\rangle
      \left\langle L \kappa \middle\vert r 0  q \kappa  \right\rangle
      \left\langle L 0 \middle\vert g_1 -1 g_2 1 \right\rangle
  [1 + (-1)^{p_1+p_2+g_1+g_2-L}]
             \\
   \nonumber
   &\times&
         W(S_1\frac{1}{2} S_2 \frac{1}{2};Iq)
 W(J_1 L I_F g_2;J_2 g_1)
     \begin{Bmatrix}
  J_1 & l_1 & S_1 \\
   L & r & q \\
   J_2 & l_2 & S_2
  \end{Bmatrix}
   	.
 \end{eqnarray}

Now using relation between vector polarization and tensor moments for the density matrix of spin 1/2 particle
\begin{eqnarray}\label{rho}
 \nonumber
 \rho &=& \frac{1}{2} \left(1+\vec{P}\cdot \vec{\sigma} \right) =\frac{1}{2}\left(
                                                                              \begin{array}{cc}
                                                                                1+P_z & P_x-iP_y \\
                                                                                P_x+iP_y & 1-P_z \\
                                                                              \end{array}
                                                                            \right)
 \\
  &=& \frac{1}{2}\left(
                   \begin{array}{cc}
                     1+\tau_{10} & \sqrt{2}\tau_{1-1} \\
                     -\sqrt{2}\tau_{11} & 1-\tau_{10} \\
                   \end{array}
                 \right),
\end{eqnarray}
 we can obtain angular correlations for $(n, \gamma )$ reactions for an arbitrary neutron polarization.
 By comparing this expression with  the density matrix for neutron polarized along direction $\vec{\sigma}(\beta ,\alpha )$
 \begin{equation}
 \label{rhoAngle}
   \rho =\left(
                      \begin{array}{cc}
                        \cos^2\frac{\beta}{2}  & \sin \frac{\beta}{2} \cos \frac{\beta}{2} e^{-i\alpha} \\
                       \sin \frac{\beta}{2} \cos \frac{\beta}{2} e^{i\alpha} & \sin^2\frac{\beta}{2} \\
                      \end{array}
                    \right),
 \end{equation}
where $\beta$ and $\alpha$ are polar and azimuthal angles for a direction of the polarization, we can relate values of $\tau_{1 \kappa}$ to the direction of neutron polarization as
\begin{eqnarray}\label{tauAngle}
 \nonumber
  \tau_{1 0} &=& \cos \beta \\
   \nonumber
  \sqrt{2}\tau_{1 1} &=& -\sin \beta e^{i\alpha} \\
  \sqrt{2}\tau_{1 -1} &=& \sin \beta e^{-i\alpha}.
\end{eqnarray}

Let write eq.(\ref{TauNgamma00}) for $q=0$
\begin{eqnarray}\label{TauNgamma00q0}
     \left\langle  T_{00} \right\rangle_{q=0}
  &=&\frac{1}{16\pi\sqrt{\pi}}\sum_{Jl  S  g p    L }\tau_{00} Y_{L 0}(\theta ,\phi)
  \frac{1}{\sqrt{(2L+1)}} \\
  \nonumber
  &\times&
  (-1)^{2(J_1+J_2+S)-2I-l_1+ l_2-g_1 }
  (2J_1+1)(2J_2+1) \\
   \nonumber
   &\times&
  \sqrt{(2l_1+1)(2l_2+1) (2g_1+1)(2g_2+1)}
   \\
   \nonumber
   &\times&
      \left\langle p_1 g_1 \alpha^{\prime} \middle\vert R^{J_1} \middle\vert S l_1 \alpha\right\rangle^*
   \left\langle p_2 g_2 \alpha^{\prime} \middle\vert R^{J_2} \middle\vert S l_2 \alpha\right\rangle
       \\
   \nonumber
   &\times&
    \left\langle L 0 \middle\vert l_1 0  l_2 0  \right\rangle
           \left\langle L 0 \middle\vert g_1 -1 g_2 1 \right\rangle
  [1 + (-1)^{p_1+p_2+g_1+g_2-L}]
             \\
   \nonumber
   &\times&
         W(J_1 S L l_2;l_1 J_2)
 W(J_1 L I_F g_2;J_2 g_1),
 \end{eqnarray}
which can be written  as
\begin{equation}\label{corrq0}
 \left\langle  T_{00} \right\rangle_{q=0} = A_0 +A_1(\vec{n}_n\cdot\vec{n}_{\gamma} ) +A_2[(\vec{n}_n\cdot\vec{n}_{\gamma} )^2-\frac{1}{3}],
\end{equation}
where $\vec{n}_n$ and $\vec{n}_{\gamma}$ unit vectors in directions of neutron and $\gamma$ momenta.

To see this, we recall that the spherical function $Y_{L 0}(\theta ,\phi)$ in eq.(\ref{TauNgamma00q0}) can be rewritten as an irreducible spherical tensor in terms of
$\vec{n}_{\gamma}$ are unit vectors in the directions of  $\gamma$ momenta. Correspondingly, in the chosen coordinate system, the directions of  neutron momenta are described by spherical functions
\begin{equation}\label{neutY}
  Y_{lm}(0,0)=\delta_{m0}\sqrt{\frac{2l+1}{4\pi}},
\end{equation}
 which can be written as irreducible spherical tensors  in terms of
$\vec{n}_{n}$ unit vectors. (For explicit expressions of spherical functions in terms of unit vectors see, for example \cite{edmonds1996angular}.)
Thus, irreducible representation of the angular dependent terms in eq.(\ref{TauNgamma00q0}) is described by a product of terms which defined as a function $\Omega_{L}$
\begin{equation}\label{OmegL}
\Omega_{L}=\tau_{00} Y_{L 0}(\theta ,\phi) \left\langle L 0 \middle\vert l_1 0  l_2 0  \right\rangle Y_{l_10}(0,0)Y_{l_20}(0,0).
\end{equation}
The parameter $L$ is bounded by neutron orbital angular momenta $l_1$ and $l_2$, and
for the case of slow neutrons ($l_1\leq 1$ and $l_2\leq 1$) can have only values of 0, 1, and 2. This leads to three possible options, correspondingly:
\begin{equation}\label{Omeg0}
\Omega_{0}=\frac{1}{(4\pi)^{3/2}}[\delta_{l_10}\delta_{l_20}-\sqrt{3}\delta_{l_11}\delta_{l_21}],
\end{equation}

\begin{equation}\label{Omeg1}
\Omega_{1}=\frac{3}{(4\pi)^{3/2}}(\vec{n}_n\cdot \vec{n}_{\gamma}),
\end{equation}
and

\begin{equation}\label{Omeg2}
\Omega_{2}=\frac{9}{8\pi }\sqrt{\frac{5}{6\pi}}\delta_{l_11}\delta_{l_21}[(\vec{n}_n\cdot \vec{n}_{\gamma})^2-\frac{1}{3}].
\end{equation}

Thus, one can write
\begin{eqnarray}\label{A0}
     A_0
  &=&\frac{1}{16\pi^2}\sum_{J l  g p}
  (-1)^{2J-2I+l+g+p } (2J+1)\\
   \nonumber
   &\times&
       | \left\langle p g \alpha^{\prime} \middle\vert R^{J} \middle\vert J l \alpha\right\rangle |^2
     ,
 \end{eqnarray}

\begin{eqnarray}\label{A1}
     A_1
  &=&\frac{1}{32\pi^2}\sum_{J g p }
   (-1)^{2(J_1-I)+I_F+1  }
  (2J_1+1)(2J_2+1) \\
   \nonumber
   &\times&
  \sqrt{(2g_1+1)(2g_2+1)}
  \left\langle 1 0 \middle\vert g_1 -1 g_2 1 \right\rangle
  [1 - (-1)^{p_1+p_2+g_1+g_2}]
   \\
   \nonumber
   &\times& \left\{
    Re(  \left\langle p_1 g_1 \alpha^{\prime} \middle\vert R^{J_1} \middle\vert J_1 0 \alpha\right\rangle^*
   \left\langle p_2 g_2 \alpha^{\prime} \middle\vert R^{J_2} \middle\vert J_1 1 \alpha\right\rangle )
         \frac{(-1)^{3J_1}}{\sqrt{(2J_1+1)}} \right.
 \\
 \nonumber
 &+&  \left.    Re (   \left\langle p_1 g_1 \alpha^{\prime} \middle\vert R^{J_1} \middle\vert J_2 1 \alpha\right\rangle^*
   \left\langle p_2 g_2 \alpha^{\prime} \middle\vert R^{J_2} \middle\vert J_2 0 \alpha\right\rangle )
                \frac{(-1)^{3J_2}}{\sqrt{(2J_2+1)}}
   \right\} \\
  \nonumber
 &\times&  W(g_1g_2J_1J_2;1 I_F),
 \end{eqnarray}

and

\begin{eqnarray}\label{A_2}
     A_2
  &=&\frac{3\sqrt{6}}{64\pi^2}\sum_{J S g p}
   (-1)^{2(J_1-I)+3S+I_F }
  (2J_1+1)(2J_2+1) \\
   \nonumber
   &\times&
  \sqrt{ (2g_1+1)(2g_2+1)}
   \\
   \nonumber
   &\times&
    Re(  \left\langle p_1 g_1 \alpha^{\prime} \middle\vert R^{J_1} \middle\vert S 1 \alpha\right\rangle^*
   \left\langle p_2 g_2 \alpha^{\prime} \middle\vert R^{J_2} \middle\vert S 1 \alpha\right\rangle )
       \\
   \nonumber
   &\times&
               \left\langle 2 0 \middle\vert g_1 -1 g_2 1 \right\rangle
  [1 + (-1)^{p_1+p_2+g_1+g_2}]
             \\
   \nonumber
   &\times&
         W(1 1 J_1 J_2;2 S)
 W(g_1 g_2 J_1 J_2;2 I_F).
 \end{eqnarray}

Let us consider the case with polarized neutrons. Then, for $q=1$ and $\kappa =0, \pm 1$

 \begin{eqnarray}\label{TauNgamma00q1}
     \left\langle  T_{00} \right\rangle_{q=1}
  &=&\frac{1}{8\pi\sqrt{2\pi}}\sum_{Jl  S  g p  r  \kappa  L }\tau_{1 \kappa} Y_{L -\kappa}(\theta ,\phi) \sqrt{\frac{3  (2r+1)}{(2L+1)}}\\
  \nonumber
  &\times&
  (-1)^{2(J_1+J_2)+S_2-I+3/2-l_1+ l_2 +\kappa-g_1}
  (2J_1+1)(2J_2+1) \\
   \nonumber
   &\times&
  \sqrt{(2l_1+1)(2l_2+1) (2S_1+1)(2S_2+1)(2g_1+1)(2g_2+1)}
   \\
   \nonumber
   &\times&
      \left\langle p_1 g_1 \alpha^{\prime} \middle\vert R^{J_1} \middle\vert S_1 l_1 \alpha\right\rangle^*
   \left\langle p_2 g_2 \alpha^{\prime} \middle\vert R^{J_2} \middle\vert S_2 l_2 \alpha\right\rangle
       \\
   \nonumber
   &\times&
    \left\langle r0 \middle\vert l_1 0  l_2 0  \right\rangle
      \left\langle L \kappa \middle\vert r 0  1 \kappa  \right\rangle
      \left\langle L 0 \middle\vert g_1 -1 g_2 1 \right\rangle
  [1 + (-1)^{p_1+p_2+g_1+g_2-L}]
             \\
   \nonumber
   &\times&
         W(S_1\frac{1}{2} S_2 \frac{1}{2};I1)
 W(J_1 L I_F g_2;J_2 g_1)
     \begin{Bmatrix}
  J_1 & l_1 & S_1 \\
   L & r & 1 \\
   J_2 & l_2 & S_2
  \end{Bmatrix}
   	.
 \end{eqnarray}

One can rewrite eq.(\ref{TauNgamma00q1}) in terms of angular correlations, observing that irreducible representation of the angular dependent terms in eq.(\ref{TauNgamma00q1}) is described by a product of terms defined by a function $\Psi_{rL}$
\begin{equation}\label{PsirL}
\Psi_{rL}=\sum_{\kappa}\tau_{1\kappa} Y_{L -\kappa}(\theta ,\phi)(-1)^{\kappa} \left\langle r 0 \middle\vert l_1 0  l_2 0  \right\rangle \left\langle L \kappa \middle\vert r 0  1 \kappa  \right\rangle Y_{l_10}(0,0)Y_{l_20}(0,0).
\end{equation}
Then, for the case of $r=0$, parameter $L=1$,
which gives
\begin{equation}\label{Psi01}
\Psi_{01}=\frac{1}{4\pi}\sqrt{\frac{3}{4\pi}}(\delta_{l_10}\delta_{l_20}-\sqrt{3}\delta_{l_11}\delta_{l_21})(\vec{\sigma}\cdot \vec{n}_{\gamma}).
\end{equation}
For the case of $r=1$, parameter $L$ can have three values $0$, $1$, and $2$
which leads to three possible correlations:
\begin{equation}\label{Psi10}
\Psi_{10}=\frac{1}{(4\pi )^{3/2}\sqrt{3}}(\delta_{l_10}\delta_{l_21}+\delta_{l_11}\delta_{l_20})(\vec{\sigma}\cdot \vec{n}_{n}),
\end{equation}

\begin{equation}\label{Psi11}
\Psi_{11}=-\frac{i}{(4\pi )^{3/2}}\sqrt{\frac{3}{2}}(\delta_{l_10}\delta_{l_21}+\delta_{l_11}\delta_{l_20})(\vec{\sigma}\cdot [\vec{n}_{n}\times \vec{n}_{\gamma} ]),
\end{equation}

\begin{equation}\label{Psi12}
\Psi_{12}=\frac{1}{16 \pi}\sqrt{\frac{15}{\pi}}(\delta_{l_10}\delta_{l_21}+\delta_{l_11}\delta_{l_20})[(\vec{n}_{n}\cdot \vec{n}_{\gamma})(\vec{\sigma}\cdot \vec{n}_{\gamma})-\frac{1}{3}(\vec{\sigma}\cdot \vec{n}_{n})],
\end{equation}
correspondingly.

The  largest value of $r=2$  leads to  three possible values of parameter $L$:  $1$, $2$, and $3$,
 which correspond  to  three possible correlations:

 \begin{equation}\label{Psi21}
\Psi_{21}=-\frac{9}{8\pi}\sqrt{\frac{1}{5\pi}}\delta_{l_11}\delta_{l_21}[(\vec{n}_{n}\cdot \vec{n}_{\gamma})(\vec{\sigma}\cdot \vec{n}_{n})-\frac{1}{3}(\vec{\sigma}\cdot \vec{n}_{\gamma})],
\end{equation}

 \begin{equation}\label{Psi22}
\Psi_{22}=\frac{i}{2\pi}\sqrt{\frac{15}{2\pi}}\delta_{l_11}\delta_{l_21}[(\vec{n}_{n}\cdot \vec{n}_{\gamma})(\vec{\sigma}\cdot [\vec{n}_{n}\times \vec{n}_{\gamma} ])],
\end{equation}

\begin{equation}\label{Psi23}
 \Psi_{23} = \frac{3}{5(4\pi)^{3/2}}\sqrt{\frac{7}{10 }}\delta_{l_11}\delta_{l_21}\{
   (\vec{n}_{n}\cdot \vec{n}_{\gamma})[(\vec{\sigma}\cdot \vec{n}_{\gamma})(\vec{n}_{n}\cdot \vec{n}_{\gamma})-\frac{2}{5}(\vec{\sigma}\cdot \vec{n}_{n})]-\frac{1}{5}(\vec{\sigma}\cdot \vec{n}_{\gamma})\}.
   \end{equation}

 Therefore, there are seven possible corresponding angular correlations

\begin{eqnarray}\label{corrq1}
 \nonumber
   \left\langle  T_{00} \right\rangle_{q=1} &=& B_1\vec{\sigma}\cdot [\vec{n}_n\times\vec{n}_{\gamma} ] +B_2 (\vec{\sigma}\cdot\vec{n}_n) +B_3 (\vec{\sigma}\cdot\vec{n}_\gamma) + B_4 [(\vec{n}_{n}\cdot \vec{n}_{\gamma})(\vec{\sigma}\cdot \vec{n}_{\gamma})-\frac{1}{3}(\vec{\sigma}\cdot \vec{n}_{n})] \\  \nonumber
    &+& B_5 [(\vec{n}_{n}\cdot \vec{n}_{\gamma})(\vec{\sigma}\cdot \vec{n}_{n})-\frac{1}{3}(\vec{\sigma}\cdot \vec{n}_{\gamma})]
  +B_6 [(\vec{n}_{n}\cdot \vec{n}_{\gamma})(\vec{\sigma}\cdot [\vec{n}_{n}\times \vec{n}_{\gamma} ])]  \\
      &+& B_7 \{ (\vec{n}_{n}\cdot \vec{n}_{\gamma})[(\vec{\sigma}\cdot \vec{n}_{\gamma})(\vec{n}_{n}\cdot \vec{n}_{\gamma})-\frac{2}{5}(\vec{\sigma}\cdot \vec{n}_{n})]-\frac{1}{5}(\vec{\sigma}\cdot \vec{n}_{\gamma}) \}  .
\end{eqnarray}
 One can see, that six first correlations in this expression correspond the correlations obtained in \cite{Sushkov:1985ng}, however the last correlation $B_7$ was not considered in  \cite{Sushkov:1985ng}. The reason for this is that, as we can see from eq.(\ref{TauNgamma00q1}), for the existence of $B_7$ correlation it is not enough to have only dipole ($E1$ and $M1$) transitions. Thus, observation of nonzero value of $B_7$ will be clear evidence of the existence (and importance) of electromagnetic transitions with higher multiplicities   in neutron radiative capture.

The simplified expressions for the first three $B$-coefficient can be written as

   \begin{eqnarray}\label{B1sym}
    B_1
  &=&\frac{-3}{32\pi^2}\sum_{J  S p g} (-1)^{-I+1/2 +I_F }\\
  \nonumber
  &\times&
    (2J_1+1)(2J_2+1) \sqrt{(2g_1+1)(2g_2+1)} \\
     \nonumber
   &\times&
               \left\langle 1 0 \middle\vert g_1 -1 g_2 1 \right\rangle
  [1 - (-1)^{p_1+p_2+g_1+g_2}]
  W(g_1 g_2 J_1 J_2 ;1 I_F)
             \\
   \nonumber
   &\times& \left\{ (-1)^{2S_2-2J_2}
  \sqrt{ (2S_2+1)}  W(J_1\frac{1}{2} S_2 \frac{1}{2};I1)
     W(11J_1 J_2;1S_2)  \right.
   \\
   \nonumber
   &\times& Im(
      \left\langle p_1 g_1 \alpha^{\prime} \middle\vert R^{J_1} \middle\vert J_1 0 \alpha\right\rangle^*
   \left\langle p_2 g_2 \alpha^{\prime} \middle\vert R^{J_2} \middle\vert S_2 1 \alpha\right\rangle )
       \\
       \nonumber
   &+& (-1)^{2S_1-2J_1}
  \sqrt{ (2S_1+1)} W(S_1\frac{1}{2} J_2 \frac{1}{2};I1)
       W(11J_1 J_2;1S_1)
   \\
   \nonumber
   &\times& \left. Im (
         \left\langle p_2 g_2 \alpha^{\prime} \middle\vert R^{J_2} \middle\vert J_2 0 \alpha\right\rangle ^*
         \left\langle p_1 g_1 \alpha^{\prime} \middle\vert R^{J_1} \middle\vert S_1 1 \alpha\right\rangle )
                 \right\}
   	,
 \end{eqnarray}

\begin{eqnarray}\label{B2}
     B_2
  &=&\frac{1}{4\pi^2\sqrt{2}}\sum_{J  S   p g}  (-1)^{-I+3/2} (2J+1) \\
    \nonumber
   &\times& \left\{(-1)^{S}
  \sqrt{(2S+1)}  W(J\frac{1}{2} S \frac{1}{2};I1) \right.
   \\
   \nonumber
      &\times& \left. Re (
      \left\langle p g \alpha^{\prime} \middle\vert R^{J} \middle\vert J 0 \alpha\right\rangle^*
   \left\langle p g \alpha^{\prime} \middle\vert R^{J} \middle\vert S 1 \alpha\right\rangle )
           \right\}   	,
 \end{eqnarray}

 and
 \begin{eqnarray}\label{B3}
     B_3
  &=&\frac{1}{16\pi^2\sqrt{2}}\sum_{J   p g}
  (-1)^{-I+1/2-2J_1+2J_2+I_F }
  (2J_1+1)(2J_2+1)
     \\
   \nonumber
   &\times&
    Re(  \left\langle p_1 g_1 \alpha^{\prime} \middle\vert R^{J_1} \middle\vert J_1 0 \alpha\right\rangle^*
   \left\langle p_2 g_2 \alpha^{\prime} \middle\vert R^{J_2} \middle\vert J_2 0 \alpha\right\rangle )
   \sqrt{ (2g_1+1)(2g_2+1)}
       \\
   \nonumber
   &\times&
                \left\langle 1 0 \middle\vert g_1 -1 g_2 1 \right\rangle
  [1 - (-1)^{p_1+p_2+g_1+g_2}]
               W(J_1\frac{1}{2} J_2 \frac{1}{2};I1)
 W(g_1 g_2 J_1 J_2 ;1 I_F)
        	.
 \end{eqnarray}

 The expressions for other four coefficients are

 \begin{eqnarray}\label{B4}
     B_4
  &=&\frac{9}{32{\pi }^2} \sqrt{\frac{3}{2}} \sum_{Jl  S  p g }(\delta_{l_10}\delta_{l_21}+\delta_{l_11}\delta_{l_20}) \\
  \nonumber
  &\times&
  (-1)^{2(J_1+J_2)+S_2-I+3/2-l_1+ l_2 -g_1}
  (2J_1+1)(2J_2+1) \\
   \nonumber
   &\times&
  \sqrt{ (2S_1+1)(2S_2+1)(2g_1+1)(2g_2+1)}
   \\
   \nonumber
   &\times&
      \left\langle p_1 g_1 \alpha^{\prime} \middle\vert R^{J_1} \middle\vert S_1 l_1 \alpha\right\rangle^*
   \left\langle p_2 g_2 \alpha^{\prime} \middle\vert R^{J_2} \middle\vert S_2 l_2 \alpha\right\rangle
       \\
   \nonumber
   &\times&
          \left\langle 2 0 \middle\vert g_1 -1 g_2 1 \right\rangle
  [1 + (-1)^{p_1+p_2+g_1+g_2}]
             \\
   \nonumber
   &\times&
         W(S_1\frac{1}{2} S_2 \frac{1}{2};I1)
 W(J_1 2 I_F g_2;J_2 g_1)
     \begin{Bmatrix}
  J_1 & l_1 & S_1 \\
   2 & 1 & 1 \\
   J_2 & l_2 & S_2
  \end{Bmatrix}
   	,
 \end{eqnarray}

 \begin{eqnarray}\label{B5}
     B_5
  &=&-\frac{9}{16{\pi }^2} \sqrt{\frac{1}{2}} \sum_{J  S  p g} \\
  \nonumber
  &\times&
  (-1)^{2(J_1+J_2)+S_2-I+3/2 -g_1}
  (2J_1+1)(2J_2+1) \\
   \nonumber
   &\times&
  \sqrt{ (2S_1+1)(2S_2+1)(2g_1+1)(2g_2+1)}
   \\
   \nonumber
   &\times&
      \left\langle p_1 g_1 \alpha^{\prime} \middle\vert R^{J_1} \middle\vert S_1 1 \alpha\right\rangle^*
   \left\langle p_2 g_2 \alpha^{\prime} \middle\vert R^{J_2} \middle\vert S_2 1 \alpha\right\rangle
       \\
   \nonumber
   &\times&
          \left\langle 1 0 \middle\vert g_1 -1 g_2 1 \right\rangle
  [1 - (-1)^{p_1+p_2+g_1+g_2}]
             \\
   \nonumber
   &\times&
         W(S_1\frac{1}{2} S_2 \frac{1}{2};I1)
 W(J_1 1 I_F g_2;J_2 g_1)
     \begin{Bmatrix}
  J_1 & 1 & S_1 \\
   1 & 2 & 1 \\
   J_2 & 1 & S_2
  \end{Bmatrix}
   	,
 \end{eqnarray}

  \begin{eqnarray}\label{B6}
     B_6
  &=&\frac{3i\sqrt{5}}{8{\pi }^2}  \sum_{J  S  p g} \\
  \nonumber
  &\times&
  (-1)^{2(J_1+J_2)+S_2-I+3/2 -g_1}
  (2J_1+1)(2J_2+1) \\
   \nonumber
   &\times&
  \sqrt{ (2S_1+1)(2S_2+1)(2g_1+1)(2g_2+1)}
   \\
   \nonumber
   &\times&
      \left\langle p_1 g_1 \alpha^{\prime} \middle\vert R^{J_1} \middle\vert S_1 1 \alpha\right\rangle^*
   \left\langle p_2 g_2 \alpha^{\prime} \middle\vert R^{J_2} \middle\vert S_2 1 \alpha\right\rangle
       \\
   \nonumber
   &\times&
          \left\langle 2 0 \middle\vert g_1 -1 g_2 1 \right\rangle
  [1 + (-1)^{p_1+p_2+g_1+g_2}]
             \\
   \nonumber
   &\times&
         W(S_1\frac{1}{2} S_2 \frac{1}{2};I1)
 W(J_1 2 I_F g_2;J_2 g_1)
     \begin{Bmatrix}
  J_1 & 1 & S_1 \\
   2 & 2 & 1 \\
   J_2 & 1 & S_2
  \end{Bmatrix}
   	,
 \end{eqnarray}

 \begin{eqnarray}\label{B7}
     B_7
  &=&\frac{3\sqrt{3}}{160 {\pi }^2} \sum_{J  S  p g} \\
  \nonumber
  &\times&
  (-1)^{2(J_1+J_2)+S_2-I+3/2 -g_1}
  (2J_1+1)(2J_2+1) \\
   \nonumber
   &\times&
  \sqrt{ (2S_1+1)(2S_2+1)(2g_1+1)(2g_2+1)}
   \\
   \nonumber
   &\times&
      \left\langle p_1 g_1 \alpha^{\prime} \middle\vert R^{J_1} \middle\vert S_1 1 \alpha\right\rangle^*
   \left\langle p_2 g_2 \alpha^{\prime} \middle\vert R^{J_2} \middle\vert S_2 1 \alpha\right\rangle
       \\
   \nonumber
   &\times&
          \left\langle 3 0 \middle\vert g_1 -1 g_2 1 \right\rangle
  [1 - (-1)^{p_1+p_2+g_1+g_2}]
             \\
   \nonumber
   &\times&
         W(S_1\frac{1}{2} S_2 \frac{1}{2};I1)
 W(J_1 3 I_F g_2;J_2 g_1)
     \begin{Bmatrix}
  J_1 & 1 & S_1 \\
   3 & 2 & 1 \\
   J_2 & 1 & S_2
  \end{Bmatrix}
   	.
 \end{eqnarray}

 \section{Description of ($n-\gamma$) reactions on some nuclei}

To apply our results for analysis of recent experiments on $^{117}$Sn, $^{131}$Xe, and $^{139}$La we use expressions for the correlation coefficients of  interest calculated for nuclear spins $1/2$,  $3/2$, and $7/2$ for $E1$ and $M1$ transitions, which are presented in Appendixes \ref{App12}, \ref{App32}, and \ref{App72}.
 Since we are interested in low energy neutron induced $\gamma$ reactions in the vicinity of compound nuclear resonances, we consider only reactions dominated by compound nuclear mechanism. Therefore, we can write $\left\langle p g \middle\vert R^{J} \middle\vert S lK \right\rangle$  in the Breit-Wigner resonance approximation  as \cite{Bunakov:1982is}
\begin{eqnarray}\label{BWamp}
\left\langle p g \middle\vert R^{J} \middle\vert S lK \right\rangle
\nonumber\\
&=& \sum_K
i \frac
{\sqrt{\Gamma^{\gamma}_{l_K}(pg)}\sqrt{\Gamma^{\rm n}_{l_K}(S_K)}}
{E-E_K+i\Gamma_K/2}
e^{i\delta_{l_K}(S_K)}
\end{eqnarray}
where $E_K$, $\Gamma_K$, and $\Gamma^n_{l_K}$ are the energy,  total width, and  partial neutron width of the $K$th nuclear compound resonance, $\Gamma^{\gamma}_{l_K}(pg)$ is the width of the  $\gamma$ transition,  $E$ is the neutron energy, and $\delta_{l_K}$ is the potential scattering phase shift.
For the case of parity violation (PV),
the matrix elements for  for slow neutrons can be written  in the Breit-Wigner resonance approximation with one $s$ resonance and one $p$ resonance as \cite{Bunakov:1982is}
\begin{eqnarray}\label{BWampSP}
\left\langle p g \middle\vert R^{J} \middle\vert S l \right\rangle
\nonumber\\
&=&
 \frac
{\sqrt{\Gamma^{\gamma}_{l^{\prime}}(pg)}(-iv)\sqrt{\Gamma^{\rm n}_{l}(S)}}
{(E-E_l+i\Gamma_l/2)(E-E_{l^{\prime}}+i\Gamma_{l^{\prime}}/2)}
e^{i\delta_{l}(S)}
\end{eqnarray}
where $l\neq l^{\prime}$, and  $v$ is the matrix elements for PV mixing between
$s$- and $p$-wave compound resonances
\begin{equation}
 v= -<{\phi_s}|V_{\not{P}}|{\phi_p}>
  \label{eq:me}
\end{equation}
due to   $V_{\not{P}}$ (PV)  interactions.
It should be noted that, in general, there are other nuclear reaction mechanizms which can contribute to these matrix elements. It has been shown  \cite{Bunakov:1982is} and confirmed by many experiments (see, for ecample \cite{Mitchell2001157} and references therein) that for PV matrix elements compound resonance mixing is a dominant mechanism for PV in heavy nuclei. For parity conserving matrix elements, other mechanizms of radiative capture  (for example \cite{Lynn}, direct capture and valence capture) can contribte to these matrix elements. That can result in a background in radiative capture  amplitude and in additional interference terms. We do not include these possible contributions in eq.(\ref{BWamp}) because we are interested in a study of radiative capture in the vicinity of compound nuclear resonances in heavy nuclei, where the main contribution comes from compound resonance radiative capture. However, in general, by increasing experimental accuracy, one can be sencitive to these additional mechanisms by observing decripancies in the description of the set of  different angular correlations using only compound resonance radiative capture mechanism.

\subsection{Description of ($n-\gamma$) reactions on  $^{117}Sn$}

To apply our results for analysis of recent experiment on $^{117}$Sn one can use the calculated  correlation coefficients for a target with spin $1/2$ for $M1$ ($g=1$ and $p=0$) and $E1$ ($g=1$ and $p=1$) $\gamma$ transitions to the ground state with spin 0 presented in appendix \ref{App12}.
For the case of ($n-\gamma$) reaction $^{117}$Sn with low energy neutrons there are only two resonances which are close to thermal neutron energy \cite{Mughabghab}: the $p$-wave resonance with total spin 1 at the energy $E_p=1.327 eV$ and  two $s$-wave resonances with spin 1 at the energy $E_{s0}=-81.02 eV$ and with spin 1 at $E_{s1}=38.80 eV$. Therefore, looking for the correlations with $\gamma$ transitions from $p$-wave resonance we keep only terms with the total spin 1.
The corresponding matrix elements defined in eqs. (\ref{BWamp}) and (\ref{BWampSP})  are

\begin{equation}\label{s12amp}
\left\langle 0, 1 \middle\vert R^{1} \middle\vert 1, 0 \right\rangle
=
i \frac
{\sqrt{\Gamma^{\gamma}_{s0}(M1)}\sqrt{\Gamma^{\rm n}_{s0}}}
{E-E_{s0}+i\Gamma_{s0}/2}
e^{i\delta_{s0}} + i \frac
{\sqrt{\Gamma^{\gamma}_{s1}(M1)}\sqrt{\Gamma^{\rm n}_{s1}}}
{E-E_{s1}+i\Gamma_{s1}/2}
e^{i\delta_{s1}} ,
\end{equation}

\begin{equation}\label{p12amp}
\left\langle 1, 1 \middle\vert R^{1} \middle\vert S, 1 \right\rangle
=
i \frac
{\sqrt{\Gamma^{\gamma}_{p}(E1)}\sqrt{\Gamma^{\rm n}_{p}(S)}}
{E-E_{p}+i\Gamma_{p}/2},
\end{equation}

\begin{eqnarray}
\label{sp12amp}
\nonumber
\left\langle 1,1 \middle\vert R^{1} \middle\vert 1,0 \right\rangle
&= &\frac
{\sqrt{\Gamma^{\gamma}_{p}(E1)}(-iv_0)\sqrt{\Gamma^{\rm n}_{s0}}}
{(E-E_p+i\Gamma_p/2)(E-E_{s0}+i\Gamma_{s0}/2)}
e^{i\delta_{s0}}   \\
&+& \frac
{\sqrt{\Gamma^{\gamma}_{p}(E1)}(-iv_1)\sqrt{\Gamma^{\rm n}_{s1}}}
{(E-E_p+i\Gamma_p/2)(E-E_{s1}+i\Gamma_{s1}/2)}
e^{i\delta_{s1}} ,
\end{eqnarray}
and
\begin{eqnarray}
\label{ps12amp}
\nonumber
\left\langle 0,1 \middle\vert R^{1} \middle\vert S,1 \right\rangle
&=& \frac
{\sqrt{\Gamma^{\gamma}_{s0}(M1)}(-iv_0)\sqrt{\Gamma^{\rm n}_{p}(S)}}
{(E-E_p+i\Gamma_p/2)(E-E_{s0}+i\Gamma_{s0}/2)} \\
&+& \frac
{\sqrt{\Gamma^{\gamma}_{s1}(M1)}(-iv_1)\sqrt{\Gamma^{\rm n}_{p}(S)}}
{(E-E_p+i\Gamma_p/2)(E-E_{s1}+i\Gamma_{s1}/2)}
.
\end{eqnarray}

Therefore, we obtain
\begin{eqnarray}\label{A0Sn}
\nonumber
 A_0&=&\frac{3}{16\pi^2}\left(\frac{\Gamma^{\gamma}_{s1}(M1)\Gamma^{\rm n}_{s1}}{(E-E_{s1})^2+\Gamma_{s1}^2/4} + \frac{\Gamma^{\gamma}_{s0}(M1)\Gamma^{\rm n}_{s0}}{(E-E_{s0})^2+\Gamma_{s0}^2/4} +\frac{\Gamma^{\gamma}_{p}(E1)\Gamma^{\rm n}_{p}}{(E-E_{p})^2+\Gamma_{p}^2/4} \right. \\
 &+&  \left. 2 \frac{\sqrt{\Gamma^{\gamma}_{s1}(M1)\Gamma^{\gamma}_{s0}(M1)\Gamma^{\rm n}_{s0}\Gamma^{\rm n}_{s1}}[(E-E_{s0})(E-E_{s1})+\Gamma_{s0}\Gamma_{s1}/4]}{[(E-E_{s0})^2+\Gamma_{s0}^2/4][(E-E_{s1})^2+\Gamma_{s1}^2/4]}
   \right),
\end{eqnarray}

\begin{eqnarray}\label{A1Sn}
\nonumber
 A_1 &=&- \frac{3\sqrt{3}}{8\sqrt{2} \pi^2} \left(    \frac{\sqrt{\Gamma^{\gamma}_{s0}(M1)\Gamma^{\gamma}_{p}(E1)\Gamma^{\rm n}_{s0}\Gamma^{\rm n}_{p}(1)}[(E-E_{s0})(E-E_p)+\frac{\Gamma_{s0}\Gamma_{p}}{4}]}{[(E-E_{s0})^2+\Gamma^2_{s0}/4][(E-E_p)^2+\Gamma^2_{p}/4]}
 \right.  \\
 &+& \left. \frac{\sqrt{\Gamma^{\gamma}_{s1}(M1)\Gamma^{\gamma}_{p}(E1)\Gamma^{\rm n}_{s1}\Gamma^{\rm n}_{p}(1)}[(E-E_{s1})(E-E_p)+\frac{\Gamma_{s1}\Gamma_{p}}{4}]}{[(E-E_{s1})^2+\Gamma^2_{s1}/4][(E-E_p)^2+\Gamma^2_{p}/4]}   \right),
\end{eqnarray}

\begin{equation}\label{A2Sn}
 A_2=\frac{9}{64\pi^2}\left(\frac{\Gamma^{\gamma}_{p}(E1)[\Gamma^{\rm n}_{p}(1)-2\Gamma^{\rm n}_{p}(0)]}{(E-E_{p})^2+\Gamma_{p}^2/4}  \right),
\end{equation}

\begin{eqnarray}\label{B1Sn}
\nonumber
 B_1 &=& -\frac{-3\sqrt{6}}{32 \pi^2} \left( \left[  \frac{\sqrt{\Gamma^{\gamma}_{s1}(M1)\Gamma^{\gamma}_{p}(E1)\Gamma^{\rm n}_{s1}\Gamma^{\rm n}_{p}(1)}}{[(E-E_{s1})^2+\Gamma^2_{s1}/4][(E-E_p)^2+\Gamma^2_{p}/4]} [(E-E_{s1})\Gamma_{p}  -(E-E_p)\Gamma_{s1}]    \right] \right. \\
  \nonumber
  &-&  \sqrt{2} \left[ \frac{\sqrt{\Gamma^{\gamma}_{s1}(M1)\Gamma^{\gamma}_{p}(E1)\Gamma^{\rm n}_{s1}\Gamma^{\rm n}_{p}(0)}}{[(E-E_{s1})^2+\Gamma^2_{s1}/4][(E-E_p)^2+\Gamma^2_{p}/4]} [(E-E_{s1})\Gamma_{p}  -(E-E_p)\Gamma_{s1}]        \right] \\
   \nonumber
  &+&   \left[ \frac{\sqrt{\Gamma^{\gamma}_{s0}(M1)\Gamma^{\gamma}_{p}(E1)\Gamma^{\rm n}_{s0}\Gamma^{\rm n}_{p}(1)}}{[(E-E_{s0})^2+\Gamma^2_{s0}/4][(E-E_p)^2+\Gamma^2_{p}/4]} [(E-E_{s0})\Gamma_{p}  -(E-E_p)\Gamma_{s0}]        \right]\\
    &-& \left. \sqrt{2} \left[ \frac{\sqrt{\Gamma^{\gamma}_{s0}(M1)\Gamma^{\gamma}_{p}(E1)\Gamma^{\rm n}_{s0}\Gamma^{\rm n}_{p}(0)}}{[(E-E_{s0})^2+\Gamma^2_{s0}/4][(E-E_p)^2+\Gamma^2_{p}/4]} [(E-E_{s0})\Gamma_{p}  -(E-E_p)\Gamma_{s0}]        \right]   \right)
\end{eqnarray}

\begin{eqnarray}\label{B2Sn}
 \nonumber
 B_2 &=& \frac{-\sqrt{3}}{8 \pi^2} \left(
          \frac{\sqrt{2}v_0\sqrt{\Gamma^{\rm n}_{s0}\Gamma^{\rm n}_{p}(1)}[(E-E_p)\Gamma^{\gamma}_{s0}(M1)+(E-E_{s0})\Gamma^{\gamma}_{p}(E1) ]}{[(E-E_{s0})^2+\Gamma^2_{s0}/4][(E-E_p)^2+\Gamma^2_{p}/4]}  \right. \\
           \nonumber
          &-&  \frac{v_0\sqrt{\Gamma^{\rm n}_{s0}\Gamma^{\rm n}_{p}(0)}[(E-E_p)\Gamma^{\gamma}_{s0}(M1)+(E-E_{s0})\Gamma^{\gamma}_{p}(E1) ]}{[(E-E_{s0})^2+\Gamma^2_{s0}/4][(E-E_p)^2+\Gamma^2_{p}/4]} \\
           \nonumber
         &+&
          \frac{\sqrt{2}v_1\sqrt{\Gamma^{\rm n}_{s1}\Gamma^{\rm n}_{p}(1)}[(E-E_p)\Gamma^{\gamma}_{s1}(M1)+(E-E_{s1})\Gamma^{\gamma}_{p}(E1) ]}{[(E-E_{s1})^2+\Gamma^2_{s1}/4][(E-E_p)^2+\Gamma^2_{p}/4]}   \\
           \nonumber
          &-&  \frac{v_1\sqrt{\Gamma^{\rm n}_{s1}\Gamma^{\rm n}_{p}(0)}[(E-E_p)\Gamma^{\gamma}_{s1}(M1)+(E-E_{s1})\Gamma^{\gamma}_{p}(E1) ]}{[(E-E_{s1})^2+\Gamma^2_{s1}/4][(E-E_p)^2+\Gamma^2_{p}/4]} \\
           \nonumber
                    &+&
          \frac{\sqrt{2}v_1\sqrt{\Gamma^{\gamma}_{s1}(M1)\Gamma^{\gamma}_{s0}(M1)\Gamma^{\rm n}_{s1}\Gamma^{\rm n}_{p}(1)}[(E-E_p)(E-E_{s1})-\Gamma_{p}\Gamma_{s1}/4 ]}{[(E-E_{s0})^2+\Gamma^2_{s0}/4][(E-E_{s1})^2+\Gamma^2_{s1}/4][(E-E_p)^2+\Gamma^2_{p}/4]}   \\
           \nonumber
          &-&  \frac{v_1\sqrt{\Gamma^{\gamma}_{s1}(M1)\Gamma^{\gamma}_{s0}(M1)\Gamma^{\rm n}_{s1}\Gamma^{\rm n}_{p}(0)}[(E-E_p)(E-E_{s1})-\Gamma_{p}\Gamma_{s1}/4 ]}{[(E-E_{s0})^2+\Gamma^2_{s0}/4][(E-E_{s1})^2+\Gamma^2_{s1}/4][(E-E_p)^2+\Gamma^2_{p}/4]}  \\
           \nonumber
                     &+&
          \frac{\sqrt{2}v_0\sqrt{\Gamma^{\gamma}_{s1}(M1)\Gamma^{\gamma}_{s0}(M1)\Gamma^{\rm n}_{s0}\Gamma^{\rm n}_{p}(1)}[(E-E_p)(E-E_{s0})-\Gamma_{p}\Gamma_{s0}/4 ]}{[(E-E_{s0})^2+\Gamma^2_{s0}/4][(E-E_{s1})^2+\Gamma^2_{s1}/4][(E-E_p)^2+\Gamma^2_{p}/4]}   \\
                    &-&  \left.   \frac{v_0\sqrt{\Gamma^{\gamma}_{s1}(M1)\Gamma^{\gamma}_{s0}(M1)\Gamma^{\rm n}_{s0}\Gamma^{\rm n}_{p}(0)}[(E-E_p)(E-E_{s0})-\Gamma_{p}\Gamma_{s0}/4 ]}{[(E-E_{s0})^2+\Gamma^2_{s0}/4][(E-E_{s1})^2+\Gamma^2_{s1}/4][(E-E_p)^2+\Gamma^2_{p}/4]}
           \right),
\end{eqnarray}

\begin{eqnarray}\label{B3Sn}
 B_3 &=& -\frac{3}{8 \pi^2} \left(
          \frac{v_0\Gamma^{\rm n}_{s0}\sqrt{\Gamma^{\gamma}_{s0}(M1)\Gamma^{\gamma}_{p}(E1)}(E-E_p) }{[(E-E_{s0})^2+\Gamma^2_{s0}/4][(E-E_p)^2+\Gamma^2_{p}/4]}  \right. \\
         \nonumber
        &+&
         \frac{v_0\sqrt{\Gamma^{\gamma}_{s0}(M1)\Gamma^{\gamma}_{p}(E1)\Gamma^{\rm n}_{s0}\Gamma^{\rm n}_{s1}} }{[(E-E_{s0})^2+\Gamma^2_{s0}/4][(E-E_{s1})^2+\Gamma^2_{s1}/4][(E-E_p)^2+\Gamma^2_{p}/4]} \\
         \nonumber
     &\times&     \left[ 2(E-E_p) [(E-E_{s0}) (E-E_{s1}) +\Gamma_{s1}\Gamma_{s0}/4] +\frac{\Gamma_{p}}{2} [(E-E_{s0}) \Gamma_{s1} - (E-E_{s1})\Gamma_{s0}] \right] \\
     \nonumber
      &+&         \frac{v_1\Gamma^{\rm n}_{s1}\sqrt{\Gamma^{\gamma}_{s1}(M1)\Gamma^{\gamma}_{p}(E1)}(E-E_p) }{[(E-E_{s1})^2+\Gamma^2_{s1}/4][(E-E_p)^2+\Gamma^2_{p}/4]}   \\
         \nonumber
        &+&
         \frac{v_1\sqrt{\Gamma^{\gamma}_{s1}(M1)\Gamma^{\gamma}_{p}(E1)\Gamma^{\rm n}_{s0}\Gamma^{\rm n}_{s1}} }{[(E-E_{s0})^2+\Gamma^2_{s0}/4][(E-E_{s1})^2+\Gamma^2_{s1}/4][(E-E_p)^2+\Gamma^2_{p}/4]} \\
         \nonumber
     &\times&    \left. \left[ 2(E-E_p) [(E-E_{s0}) (E-E_{s1}) +\Gamma_{s1}\Gamma_{s0}/4] +\frac{\Gamma_{p}}{2} [(E-E_{s1}) \Gamma_{s0} - (E-E_{s0})\Gamma_{s1}] \right]
                    \right).
\end{eqnarray}

\subsection{Description of ($n-\gamma $) reactions on $^{131}Xe$}

To  analyze  recent experiments on $^{131}$Xe with spin $3/2$ , we use results of calculations presented in Appendix \ref{App32}.
For  ($n-\gamma$) reaction $^{131}$Xe with low energy neutrons there are only two resonances which are close to the thermal neutron energy\cite{Mughabghab}: $p$-wave resonance with unkown total spin  at the energy $E_p=3.2 eV$ and  two $s$-wave resonances: one with spin 2 at the energy $E_{s2}=14.410 eV$ and another one with spin 1 at $E_{s1}=-5.65 eV$. Therefore,  non-zero matrix elements with only s-resonances defined in eqs. (\ref{BWamp})  are

\begin{eqnarray}\label{s32amp}
\nonumber
 \left\langle 0, 1 \middle\vert R^{1} \middle\vert 1, 0 \right\rangle &=& i \frac
{\sqrt{\Gamma^{\gamma}_{s1}(M1)}\sqrt{\Gamma^{\rm n}_{s1}}}
{E-E_{s1}+i\Gamma_{s1}/2}
e^{i\delta_{s1}} , \\
  \left\langle 0, 1 \middle\vert R^{2} \middle\vert 2, 0 \right\rangle &=& i \frac
{\sqrt{\Gamma^{\gamma}_{s2}(M1)}\sqrt{\Gamma^{\rm n}_{s2}}}
{E-E_{s2}+i\Gamma_{s2}/2}
e^{i\delta_{s2}}.
\end{eqnarray}

The matrix elements with spin 1 p-resonance defined in eqs. (\ref{BWamp}) and (\ref{BWampSP})  are:

\begin{equation}\label{p32amp1}
\left\langle 1, 1 \middle\vert R^{1} \middle\vert S, 1 \right\rangle
=
i \frac
{\sqrt{\Gamma^{\gamma}_{p}(E1)}\sqrt{\Gamma^{\rm n}_{p}(S)}}
{E-E_{p}+i\Gamma_{p}/2},
\end{equation}

\begin{eqnarray}
\label{sp32amp1}
\nonumber
\left\langle 1,1 \middle\vert R^{1} \middle\vert 1,0 \right\rangle
&= &\frac
{\sqrt{\Gamma^{\gamma}_{p}(E1)}(-iv_1)\sqrt{\Gamma^{\rm n}_{s1}}}
{(E-E_p+i\Gamma_p/2)(E-E_{s1}+i\Gamma_{s1}/2)}
e^{i\delta_{s1}} ,  \\
\left\langle 0,1 \middle\vert R^{1} \middle\vert S,1 \right\rangle
&=& \frac
{\sqrt{\Gamma^{\gamma}_{s1}(M1)}(-iv_1)\sqrt{\Gamma^{\rm n}_{p}(S)}}
{(E-E_p+i\Gamma_p/2)(E-E_{s1}+i\Gamma_{s1}/2)},
\end{eqnarray}

and for spin 2 p-resonance:

\begin{equation}\label{p32amp2}
\left\langle 1, 1 \middle\vert R^{2} \middle\vert S, 1 \right\rangle
=
i \frac
{\sqrt{\Gamma^{\gamma}_{p}(E1)}\sqrt{\Gamma^{\rm n}_{p}(S)}}
{E-E_{p}+i\Gamma_{p}/2},
\end{equation}

\begin{eqnarray}
\label{sp32amp2}
\nonumber
\left\langle 1,1 \middle\vert R^{2} \middle\vert 2,0 \right\rangle
&= &\frac
{\sqrt{\Gamma^{\gamma}_{p}(E1)}(-iv_2)\sqrt{\Gamma^{\rm n}_{s2}}}
{(E-E_p+i\Gamma_p/2)(E-E_{s2}+i\Gamma_{s2}/2)}
e^{i\delta_{s2}} ,  \\
\left\langle 0,1 \middle\vert R^{2} \middle\vert S,1 \right\rangle
&=& \frac
{\sqrt{\Gamma^{\gamma}_{s1}(M1)}(-iv_2)\sqrt{\Gamma^{\rm n}_{p}(S)}}
{(E-E_p+i\Gamma_p/2)(E-E_{s2}+i\Gamma_{s2}/2)}.
\end{eqnarray}

Therefore, for spin 1 p-wave resonance we obtain
\begin{equation}\label{A0Xe1}
 A^{J=1}_0=\frac{1}{16\pi^2}\left(3\frac{\Gamma^{\gamma}_{s1}(M1)\Gamma^{\rm n}_{s1}}{(E-E_{s1})^2+\Gamma_{s1}^2/4} + 5\frac{\Gamma^{\gamma}_{s2}(M1)\Gamma^{\rm n}_{s2}}{(E-E_{s2})^2+\Gamma_{s2}^2/4}
 + 3\frac{\Gamma^{\gamma}_{p}(E1)\Gamma^{\rm n}_{p}(1)}{(E-E_{p})^2+\Gamma_{p}^2/4}
   \right),
\end{equation}

\begin{equation}\label{A1Xe1}
 A^{J=1}_1 = -\frac{3\sqrt{3}}{8\sqrt{2} \pi^2} \left(    \frac{\sqrt{\Gamma^{\gamma}_{s1}(M1)\Gamma^{\gamma}_{p}(E1)\Gamma^{\rm n}_{s1}\Gamma^{\rm n}_{p}(1)}[(E-E_{s1})(E-E_p)+\frac{\Gamma_{s1}\Gamma_{p}}{4}]}{[(E-E_{s1})^2+\Gamma^2_{s1}/4][(E-E_p)^2+\Gamma^2_{p}/4]}   \right),
\end{equation}

\begin{equation}\label{A2Xe1}
 A^{J=1}_2=\frac{9}{320\pi^2}\left(\frac{\Gamma^{\gamma}_{p}(E1)[5\Gamma^{\rm n}_{p}(1)-\Gamma^{\rm n}_{p}(2)]}{(E-E_{p})^2+\Gamma_{p}^2/4}  \right),
\end{equation}

\begin{eqnarray}\label{B1Xe1}
\nonumber
 B^{J=1}_1 &=& -\frac{3}{32 \pi^2}\sqrt{\frac{3}{2}} \left( \left[  \frac{\sqrt{\Gamma^{\gamma}_{s1}(M1)\Gamma^{\gamma}_{p}(E1)\Gamma^{\rm n}_{s1}\Gamma^{\rm n}_{p}(1)}}{[(E-E_{s1})^2+\Gamma^2_{s1}/4][(E-E_p)^2+\Gamma^2_{p}/4]} [(E-E_{s1})\Gamma_{p}  -(E-E_p)\Gamma_{s1}]    \right] \right. \\
    &+& \left. \sqrt{5} \left[ \frac{\sqrt{\Gamma^{\gamma}_{s1}(M1)\Gamma^{\gamma}_{p}(E1)\Gamma^{\rm n}_{s1}\Gamma^{\rm n}_{p}(2)}}{[(E-E_{s1})^2+\Gamma^2_{s1}/4][(E-E_p)^2+\Gamma^2_{p}/4]} [(E-E_{s1})\Gamma_{p}  -(E-E_p)\Gamma_{s1}]        \right]   \right) ,
\end{eqnarray}

\begin{eqnarray}\label{B2Xe1}
 \nonumber
 B^{J=1}_2 &=& \frac{1}{16\sqrt{2} \pi^2} \left(
          \frac{\sqrt{15}v_1\sqrt{\Gamma^{\rm n}_{s1}\Gamma^{\rm n}_{p}(2)}[(E-E_p)\Gamma^{\gamma}_{s1}(M1)+(E-E_{s1})\Gamma^{\gamma}_{p}(E1) ]}{[(E-E_{s1})^2+\Gamma^2_{s1}/4][(E-E_p)^2+\Gamma^2_{p}/4]} \right.  \\
  &-& \left. \frac{\sqrt{3}v_1\sqrt{\Gamma^{\rm n}_{s1}\Gamma^{\rm n}_{p}(1)}[(E-E_p)\Gamma^{\gamma}_{s1}(M1)+(E-E_{s1})\Gamma^{\gamma}_{p}(E1) ]}{[(E-E_{s1})^2+\Gamma^2_{s1}/4][(E-E_p)^2+\Gamma^2_{p}/4]}
                    \right),
\end{eqnarray}

\begin{equation}\label{B3Xe1}
 B^{J=1}_3 = \frac{3}{16 \pi^2} \left(     \frac{v_1\Gamma^{\rm n}_{s1}\sqrt{\Gamma^{\gamma}_{s1}(M1)\Gamma^{\gamma}_{p}(E1)}(E-E_p) }{[(E-E_{s1})^2+\Gamma^2_{s1}/4][(E-E_p)^2+\Gamma^2_{p}/4]}   \right).
\end{equation}

For spin 2 p-wave resonance the corresponding correlations are:
\begin{equation}\label{A0Xe2}
 A^{J=2}_0=\frac{1}{16\pi^2}\left(3\frac{\Gamma^{\gamma}_{s1}(M1)\Gamma^{\rm n}_{s1}}{(E-E_{s1})^2+\Gamma_{s1}^2/4} + 5\frac{\Gamma^{\gamma}_{s2}(M1)\Gamma^{\rm n}_{s2}}{(E-E_{s2})^2+\Gamma_{s2}^2/4}
 + 5\frac{\Gamma^{\gamma}_{p}(E1)\Gamma^{\rm n}_{p}(2)}{(E-E_{p})^2+\Gamma_{p}^2/4}
   \right),
\end{equation}

\begin{eqnarray}\label{A1Xe2}
\nonumber
 A^{J=2}_1 &=& -\frac{1}{16\sqrt{2} \pi^2} \left( - 3 \sqrt{15} \frac{\sqrt{\Gamma^{\gamma}_{s1}(M1)\Gamma^{\gamma}_{p}(E1)\Gamma^{\rm n}_{s1}\Gamma^{\rm n}_{p}(1)}[(E-E_{s1})(E-E_p)+\frac{\Gamma_{s1}\Gamma_{p}}{4}]}{[(E-E_{s1})^2+\Gamma^2_{s1}/4][(E-E_p)^2+\Gamma^2_{p}/4]} \right. \\
  &+& \left. 5\frac{\sqrt{\Gamma^{\gamma}_{s2}(M1)\Gamma^{\gamma}_{p}(E1)\Gamma^{\rm n}_{s2}\Gamma^{\rm n}_{p}(2)}[(E-E_{s2})(E-E_p)+\frac{\Gamma_{s2}\Gamma_{p}}{4}]}{[(E-E_{s2})^2+\Gamma^2_{s2}/4][(E-E_p)^2+\Gamma^2_{p}/4]}
  \right),
\end{eqnarray}

\begin{equation}\label{A2Xe2}
 A^{J=2}_2=\frac{21}{128\pi^2}\left(\frac{\Gamma^{\gamma}_{p}(E1)[\Gamma^{\rm n}_{p}(1)-\Gamma^{\rm n}_{p}(2)]}{(E-E_{p})^2+\Gamma_{p}^2/4}  \right),
\end{equation}

\begin{eqnarray}\label{B1Xe2}
\nonumber
 B^{J=2}_1 &=& \frac{1}{64\sqrt{2} \pi^2} \left(9\sqrt{15} \left[  \frac{\sqrt{\Gamma^{\gamma}_{s1}(M1)\Gamma^{\gamma}_{p}(E1)\Gamma^{\rm n}_{s1}\Gamma^{\rm n}_{p}(2)}}{[(E-E_{s1})^2+\Gamma^2_{s1}/4][(E-E_p)^2+\Gamma^2_{p}/4]} [(E-E_{s1})\Gamma_{p}  -(E-E_p)\Gamma_{s1}]    \right] \right. \\
 \nonumber
 &-& 15 \left[  \frac{\sqrt{\Gamma^{\gamma}_{s2}(M1)\Gamma^{\gamma}_{p}(E1)\Gamma^{\rm n}_{s2}\Gamma^{\rm n}_{p}(1)}}{[(E-E_{s2})^2+\Gamma^2_{s2}/4][(E-E_p)^2+\Gamma^2_{p}/4]} [(E-E_{s2})\Gamma_{p}  -(E-E_p)\Gamma_{s2}]    \right] \\
 &+& \left. 5 \left[ \frac{\sqrt{\Gamma^{\gamma}_{s2}(M1)\Gamma^{\gamma}_{p}(E1)\Gamma^{\rm n}_{s2}\Gamma^{\rm n}_{p}(2)}}{[(E-E_{s2})^2+\Gamma^2_{s2}/4][(E-E_p)^2+\Gamma^2_{p}/4]} [(E-E_{s2})\Gamma_{p}  -(E-E_p)\Gamma_{s2}]        \right]   \right) ,
\end{eqnarray}

\begin{eqnarray}\label{B2Xe2}
 \nonumber
 B^{J=2}_2 &=& \frac{5}{16\sqrt{2} \pi^2} \left(
          \frac{v_2\sqrt{\Gamma^{\rm n}_{s2}\Gamma^{\rm n}_{p}(2)}[(E-E_p)\Gamma^{\gamma}_{s2}(M1)+(E-E_{s2})\Gamma^{\gamma}_{p}(E1) ]}{[(E-E_{s2})^2+\Gamma^2_{s2}/4][(E-E_p)^2+\Gamma^2_{p}/4]} \right.  \\
  &-& \left. \frac{v_2\sqrt{\Gamma^{\rm n}_{s2}\Gamma^{\rm n}_{p}(1)}[(E-E_p)\Gamma^{\gamma}_{s2}(M1)+(E-E_{s2})\Gamma^{\gamma}_{p}(E1) ]}{[(E-E_{s2})^2+\Gamma^2_{s2}/4][(E-E_p)^2+\Gamma^2_{p}/4]}
                    \right),
\end{eqnarray}

\begin{eqnarray}\label{B3Xe2}
\nonumber
 B^{J=2}_3 &=& -\frac{1}{32 \pi^2} \left(  5   \frac{v_2\Gamma^{\rm n}_{s2}\sqrt{\Gamma^{\gamma}_{s2}(M1)\Gamma^{\gamma}_{p}(E1)}(E-E_p) }{[(E-E_{s2})^2+\Gamma^2_{s2}/4][(E-E_p)^2+\Gamma^2_{p}/4]} \right. \\
 \nonumber
 &+& 3\sqrt{15}\frac{v_2\sqrt{\Gamma^{\rm n}_{s1}\Gamma^{\rm n}_{s2}\Gamma^{\gamma}_{s2}(M1)\Gamma^{\gamma}_{p}(E1)}}{[(E-E_{s1})^2+\Gamma^2_{s1}/4][(E-E_{s2})^2+\Gamma^2_{s2}/4][(E-E_p)^2+\Gamma^2_{p}/4]}
 \\
 \nonumber
 &\times &  [ (E-E_{p})(E-E_{s1})(E-E_{s2})+(E-E_{p})\Gamma_{s1}\Gamma_{s2}/4 \\
 &+& \left.
 (E-E_{s2})\Gamma_{s1}\Gamma_{p}/4-(E-E_{s1})\Gamma_{p}\Gamma_{s2}/4   ]
    \right).
\end{eqnarray}

\subsection{Description of ($n-\gamma $) reactions on $^{139}La$}

Let us apply  results of calculations of the angular correlations for spin 7/2 in Appendix \ref{App72} for $^{139}$La.
Then, for the case of the ($n-\gamma$) reaction  with low energy neutrons there are only two resonances which are close to the thermal neutron energy\cite{Mughabghab}: the $p$-wave resonance with total spin 4 at the energy $E_p=0.734 eV$ and  two $s$-wave resonances with spin 4 at the energy $E_{s0}=-48.63 eV$ and with spin 3 at $E_{s1}=72.3 eV$. Therefore, looking for the correlations with $\gamma$ transitions from the $p$-wave resonance we keep only terms with the total spin 4, except the $A_0$ coefficient which contains a ``background'' from both $s$-wave resonances.
The corresponding matrix elements defined in eqs. (\ref{BWamp}) and (\ref{BWampSP})  are:

\begin{equation}\label{s0amp}
\left\langle 1, 1 \middle\vert R^{4} \middle\vert 4, 0 \right\rangle
=
i \frac
{\sqrt{\Gamma^{\gamma}_{s0}(E1)}\sqrt{\Gamma^{\rm n}_{s0}}}
{E-E_{s0}+i\Gamma_{s0}/2}
e^{i\delta_{s0}},
\end{equation}

\begin{equation}\label{s1amp}
\left\langle 1, 1 \middle\vert R^{3} \middle\vert 3, 0 \right\rangle
=
i \frac
{\sqrt{\Gamma^{\gamma}_{s1}(E1)}\sqrt{\Gamma^{\rm n}_{s1}}}
{E-E_{s1}+i\Gamma_{s1}/2}
e^{i\delta_{s1}},
\end{equation}

\begin{equation}\label{pamp}
\left\langle 0, 1 \middle\vert R^{4} \middle\vert S, 1 \right\rangle
=
i \frac
{\sqrt{\Gamma^{\gamma}_{p}(M1)}\sqrt{\Gamma^{\rm n}_{p}(S)}}
{E-E_{p}+i\Gamma_{p}/2},
\end{equation}

\begin{equation}
\label{spamp}
\left\langle 0,1 \middle\vert R^{4} \middle\vert 4,0 \right\rangle
= \frac
{\sqrt{\Gamma^{\gamma}_{p}(M1)}(-iv)\sqrt{\Gamma^{\rm n}_{s0}}}
{(E-E_p+i\Gamma_p/2)(E-E_{s0}+i\Gamma_{s0}/2)}
e^{i\delta_{s0}},
\end{equation}
and
\begin{equation}
\label{psamp}
\left\langle 1,1 \middle\vert R^{4} \middle\vert S,1 \right\rangle
= \frac
{\sqrt{\Gamma^{\gamma}_{s0}(E1)}(-iv)\sqrt{\Gamma^{\rm n}_{p}(S)}}
{(E-E_p+i\Gamma_p/2)(E-E_{s0}+i\Gamma_{s0}/2)}
.
\end{equation}

Therefore, we obtain
\begin{eqnarray}\label{A0La}
\nonumber
 A_0&=&\frac{1}{16\pi^2}\left(7\frac{\Gamma^{\gamma}_{s1}(E1)\Gamma^{\rm n}_{s1}}{(E-E_{s1})^2+\Gamma_{s1}^2/4} + 9\frac{\Gamma^{\gamma}_{s0}(E1)\Gamma^{\rm n}_{s0}}{(E-E_{s0})^2+\Gamma_{s0}^2/4} \right. \\
 &+& \left. 9\frac{\Gamma^{\gamma}_{p}(M1)\Gamma^{\rm n}_{p}}{(E-E_{p})^2+\Gamma_{p}^2/4} \right),
\end{eqnarray}

\begin{eqnarray}\label{A1La}
\nonumber
 A_1 &=&- \frac{9}{64 \pi^2} \left(3 \left[   \frac{\sqrt{\Gamma^{\gamma}_{s1}(E1)\Gamma^{\gamma}_{p}(M1)\Gamma^{\rm n}_{s1}\Gamma^{\rm n}_{p}(3)}}{[(E-E_{s1})^2+\Gamma^2_{s1}/4][(E-E_p)^2+\Gamma^2_{p}/4]} [(E-E_{s1})(E-E_p)+\frac{\Gamma_{s1}\Gamma_{p}}{4}]    \right] \right. \\
  &+& \left. 2\sqrt{15} \left[ \frac{\sqrt{\Gamma^{\gamma}_{s0}(E1)\Gamma^{\gamma}_{p}(M1)\Gamma^{\rm n}_{s0}\Gamma^{\rm n}_{p}(4)}}{[(E-E_{s0})^2+\Gamma^2_{s0}/4][(E-E_p)^2+\Gamma^2_{p}/4]} [(E-E_{s0})(E-E_p)+\frac{\Gamma_{s0}\Gamma_{p}}{4}]        \right]   \right),
\end{eqnarray}

\begin{equation}\label{A2La}
 A_2=\frac{297}{8960\pi^2}\left(\frac{\Gamma^{\gamma}_{p}(M1)[7\Gamma^{\rm n}_{p}(4)-5\Gamma^{\rm n}_{p}(3)]}{(E-E_{p})^2+\Gamma_{p}^2/4}  \right),
\end{equation}

\begin{eqnarray}\label{B1La}
\nonumber
 B_1 &=& \frac{9}{256 \pi^2} \left(9 \left[  \frac{\sqrt{\Gamma^{\gamma}_{s1}(E1)\Gamma^{\gamma}_{p}(M1)\Gamma^{\rm n}_{s1}\Gamma^{\rm n}_{p}(3)}}{[(E-E_{s1})^2+\Gamma^2_{s1}/4][(E-E_p)^2+\Gamma^2_{p}/4]} [(E-E_{s1})\Gamma_{p}  -(E-E_p)\Gamma_{s1}]    \right] \right. \\
  \nonumber
  &-&  3\sqrt{35} \left[ \frac{\sqrt{\Gamma^{\gamma}_{s1}(E1)\Gamma^{\gamma}_{p}(M1)\Gamma^{\rm n}_{s1}\Gamma^{\rm n}_{p}(4)}}{[(E-E_{s1})^2+\Gamma^2_{s1}/4][(E-E_p)^2+\Gamma^2_{p}/4]} [(E-E_{s1})\Gamma_{p}  -(E-E_p)\Gamma_{s1}]        \right] \\
   \nonumber
  &+&  5\sqrt{21} \left[ \frac{\sqrt{\Gamma^{\gamma}_{s0}(E1)\Gamma^{\gamma}_{p}(M1)\Gamma^{\rm n}_{s0}\Gamma^{\rm n}_{p}(3)}}{[(E-E_{s0})^2+\Gamma^2_{s0}/4][(E-E_p)^2+\Gamma^2_{p}/4]} [(E-E_{s0})\Gamma_{p}  -(E-E_p)\Gamma_{s0}]        \right]\\
    &-& \left. \sqrt{15} \left[ \frac{\sqrt{\Gamma^{\gamma}_{s0}(E1)\Gamma^{\gamma}_{p}(M1)\Gamma^{\rm n}_{s0}\Gamma^{\rm n}_{p}(4)}}{[(E-E_{s0})^2+\Gamma^2_{s0}/4][(E-E_p)^2+\Gamma^2_{p}/4]} [(E-E_{s0})\Gamma_{p}  -(E-E_p)\Gamma_{s0}]        \right]   \right)
\end{eqnarray}

\begin{eqnarray}\label{B2La}
 \nonumber
 B_2 &=& \frac{-3\sqrt{3}}{16 \pi^2} \left(
          \frac{\sqrt{5}v\sqrt{\Gamma^{\rm n}_{s0}\Gamma^{\rm n}_{p}(4)}[(E-E_p)\Gamma^{\gamma}_{s0}(E1)+(E-E_{s0})\Gamma^{\gamma}_{p}(M1) ]}{[(E-E_{s0})^2+\Gamma^2_{s0}/4][(E-E_p)^2+\Gamma^2_{p}/4]}  \right. \\
          &-& \left. \frac{\sqrt{7}v\sqrt{\Gamma^{\rm n}_{s0}\Gamma^{\rm n}_{p}(3)}[(E-E_p)\Gamma^{\gamma}_{s0}(E1)+(E-E_{s0})\Gamma^{\gamma}_{p}(M1) ]}{[(E-E_{s0})^2+\Gamma^2_{s0}/4][(E-E_p)^2+\Gamma^2_{p}/4]} \right),
\end{eqnarray}

\begin{eqnarray}\label{B3La}
 B_3 &=& -\frac{1}{128 \pi^2} \left(
         90 \frac{v\Gamma^{\rm n}_{s0}\sqrt{\Gamma^{\gamma}_{s0}(E1)\Gamma^{\gamma}_{p}(M1)}(E-E_p) }{[(E-E_{s0})^2+\Gamma^2_{s0}/4][(E-E_p)^2+\Gamma^2_{p}/4]}  \right. \\
         \nonumber
        &+&  9\sqrt{21}
         \frac{v\sqrt{\Gamma^{\gamma}_{s0}(E1)\Gamma^{\gamma}_{p}(M1)\Gamma^{\rm n}_{s0}\Gamma^{\rm n}_{s1}} }{[(E-E_{s0})^2+\Gamma^2_{s0}/4][(E-E_{s1})^2+\Gamma^2_{s1}/4][(E-E_p)^2+\Gamma^2_{p}/4]} \\
         \nonumber
     &\times&    \left. \left[ 2(E-E_p) [(E-E_{s0}) (E-E_{s1}) +\Gamma_{s1}\Gamma_{s0}/4] +\frac{\Gamma_{p}}{2} [(E-E_{s0}) \Gamma_{s1} - (E-E_{s1})\Gamma_{s0}] \right]
                    \right).
\end{eqnarray}

\section{Conclusions}
The calculations of angular correlations for low energy n-$\gamma$ reactions for dipole transitions obtained in nuclear reaction formalism confirm results of previous calculations \cite{Sushkov:1985ng}. This gives the assurance for the correct extraction of spin-dependent neutron partial widths from the recently obtained \cite{Okudaira:2017dun,Yamamoto:2020xtz,Okudaira:2021svc,Koga:2022tb,Endo:2022nxr,Okudaira:2022wpa,Okuizumi:2024rkp} and future experimental data.    Using nuclear theory formalism we presented a multi resonance approach for calculation of angular correlations with arbitrary multiplicity and provided  specific expressions for the correlation with dipole transitions in the vicinity of the p-wave resonance for $^{117}$Sn,  $^{139}$La and $^{131}$Xe  nuclei. We observed that the correlation $B_7$ has non-zero value only if n-$\gamma$ reactions have electromagnetic transitions with multiplicities larger than 1. Therefore, the measurement of $B_7$ could be an unique test of the existence and importance of higher multiplicity  electromagnetic transitions  radiation neutron capture. Also we summarized and presented in sec. III the reason for the possible discrepancies in different calculations of n-$\gamma$ correlations.


\begin{acknowledgments}
This material is based on work supported by the U.S. Department of Energy Office of Science, Office of Nuclear Physics program under Award No. DE-SC0020687. The authors would like to thank Dr. Jonathan Curole for checking the  signs of some calculated angular correlations.
\end{acknowledgments}

\appendix
\section{Description of ($n-\gamma $) reactions on nuclei with  spin $I=1/2$}
\label{App12}

Let us calculate the measured correlation coefficients for a target with spin $1/2$ for $M1$ ($g=1$ and $p=0$) and $E1$ ($g=1$ and $p=1$) $\gamma$ transitions to the ground state with spin 0. Thus

\begin{eqnarray}\label{A012}
\nonumber
 A_0&=& \frac{1}{16 \pi ^2}\left( \middle\vert \left\langle 0, 1 \middle\vert R^{0}\middle\vert 0,0 \right\rangle  \middle\vert^2+3 |\left\langle 0, 1 \middle\vert R^{1}\middle\vert 1,0 \right\rangle |^2 \right. \\
 &+& \left. \middle\vert \left\langle 1, 1 \middle\vert R^{0}\middle\vert 0,1 \right\rangle  \middle\vert^2+3 |\left\langle 1, 1 \middle\vert R^{1}\middle\vert 1,1 \right\rangle |^2  \right),
\end{eqnarray}

\begin{equation}\label{A112}
\begin{split}
A_1&=-\frac{3\sqrt{3}}{16 \sqrt{2} \pi ^2} \Bigl([ \left\langle 1, 1 \middle\vert R^{1}\middle\vert 1,1 \right\rangle ^*\left\langle 0 , 1 \middle\vert R^{1}\middle\vert 1,0 \right\rangle +\left\langle 0, 1 \middle\vert R^{1}\middle\vert 1,0 \right\rangle ^* \left\langle 1, 1 \middle\vert R^{1}\middle\vert 1,1 \right\rangle]  \\
&+[ \left\langle 0, 1 \middle\vert R^{1}\middle\vert 1,1 \right\rangle ^*\left\langle 1 ,1 \middle\vert R^{1}\middle\vert 1,0 \right\rangle +\left\langle 1, 1 \middle\vert R^{1}\middle\vert 1,0 \right\rangle ^*\left\langle 0 ,1 \middle\vert R^{1}\middle\vert 1,1 \right\rangle ]\Bigr) ,
 \end{split}
\end{equation}

\begin{equation}\label{A212}
\begin{split}
A_2 &=\frac{9}{64 \pi ^2} \Bigl( |\left\langle 0, 1 \middle\vert R^{1}\middle\vert 1,1 \right\rangle |^2 + |\left\langle 1, 1 \middle\vert R^{1}\middle\vert 1,1 \right\rangle |^2 \\
&-2 |\left\langle 0, 1 \middle\vert R^{1}\middle\vert 0,1 \right\rangle |^2 - 2 |\left\langle 1, 1 \middle\vert R^{1}\middle\vert 0,1 \right\rangle |^2
 \Bigr) ,
 \end{split}
\end{equation}

\begin{equation}\label{B112}
\begin{split}
B_1 & = -\frac{3i\sqrt{3}}{32 \pi ^2}  \Bigl( 2 \bigl( \left\langle 1, 1 \middle\vert R^{1}\middle\vert 1,0 \right\rangle ^*\left\langle 0,1 \middle\vert R^{1}\middle\vert 0,1 \right\rangle- \left\langle 0, 1 \middle\vert R^{1}\middle\vert 0,1 \right\rangle ^*\left\langle 1 , 1 \middle\vert R^{1}\middle\vert 1,0 \right\rangle \bigr)\\
&+ \sqrt{2} \bigl(\left\langle 1, 1 \middle\vert R^{1}\middle\vert 1,1 \right\rangle ^*\left\langle 0,1 \middle\vert R^{1}\middle\vert 1,0 \right\rangle -\left\langle 0, 1 \middle\vert R^{1}\middle\vert 1,0 \right\rangle ^*\left\langle 1,1 \middle\vert R^{1}\middle\vert 1,1 \right\rangle \bigr)\\
 &-2 \bigl( \left\langle 1, 1 \middle\vert R^{1}\middle\vert 0,1 \right\rangle ^*\left\langle 0 ,1 \middle\vert R^{1}\middle\vert 1,0 \right\rangle -\left\langle 0, 1 \middle\vert R^{1}\middle\vert 1,0 \right\rangle ^*\left\langle 1 ,1 \middle\vert R^{1}\middle\vert 0,1 \right\rangle \bigr) \\
 &+\sqrt{2}\bigl(\left\langle 0, 1 \middle\vert R^{1}\middle\vert 1,1 \right\rangle ^*\left\langle 1 ,1 \middle\vert R^{1}\middle\vert 1,0 \right\rangle -\left\langle 1, 1 \middle\vert R^{1}\middle\vert 1,0 \right\rangle ^*\left\langle 0 ,1 \middle\vert R^{1}\middle\vert 1,1 \right\rangle  \bigr)  \Bigr) ,
 \end{split}
\end{equation}

\begin{equation}\label{B212}
\begin{split}
B_2 &= \frac{1}{16 \pi ^2} \Bigl(\bigl(\left\langle 0 , 1 \middle\vert R^{0}\middle\vert 1,1 \right\rangle ^*\left\langle 0,1 \middle\vert R^{0}\middle\vert 0,0 \right\rangle +\left\langle 0,1 \middle\vert R^{0}\middle\vert 0,0 \right\rangle ^*\left\langle 0 , 1 \middle\vert R^{0}\middle\vert 1,1 \right\rangle \bigr) \\
&-\sqrt{3}\bigl(\left\langle 0 , 1 \middle\vert R^{1}\middle\vert 1,0 \right\rangle ^*\left\langle 0,1 \middle\vert R^{1}\middle\vert 0,1 \right\rangle +\left\langle 0,1 \middle\vert R^{1}\middle\vert 0,1 \right\rangle ^*\left\langle 0 , 1 \middle\vert R^{1}\middle\vert 1,0 \right\rangle \bigr) \\
 &+\sqrt{6 }\bigl( \left\langle 0, 1 \middle\vert R^{1}\middle\vert 1,1 \right\rangle ^*\left\langle 0 , 1 \middle\vert R^{1}\middle\vert 1,0 \right\rangle +\left\langle 0 , 1 \middle\vert R^{1}\middle\vert 1,0 \right\rangle ^*\left\langle 0, 1 \middle\vert R^{1}\middle\vert 1,1 \right\rangle \bigr) \\
    &+\bigl( \left\langle 1, 1 \middle\vert R^{0}\middle\vert 1,1 \right\rangle ^*\left\langle 1 , 1 \middle\vert R^{0}\middle\vert 0,0 \right\rangle +\left\langle 1 , 1 \middle\vert R^{0}\middle\vert 0,0 \right\rangle ^*\left\langle 1, 1 \middle\vert R^{0}\middle\vert 1,1 \right\rangle \bigr) \\
      &-\sqrt{3}  \bigl( \left\langle 1 ,1 \middle\vert R^{1}\middle\vert 1,0 \right\rangle ^*\left\langle 1, 1 \middle\vert R^{1}\middle\vert 0,1 \right\rangle +\left\langle 1, 1 \middle\vert R^{1}\middle\vert 0,1 \right\rangle ^*\left\langle 1 ,1 \middle\vert R^{1}\middle\vert 1,0 \right\rangle \bigr) \\
          &+ \sqrt{6} \bigl( \left\langle 1 ,1 \middle\vert R^{1}\middle\vert 1,1 \right\rangle ^*\left\langle 1 ,1 \middle\vert R^{1}\middle\vert 1,0 \right\rangle +\left\langle 1 ,1 \middle\vert R^{1}\middle\vert 1,0 \right\rangle ^*\left\langle 1 ,1 \middle\vert R^{1}\middle\vert 1,1 \right\rangle \bigr) \Bigr),
 \end{split}
\end{equation}

\begin{equation}\label{B312}
\begin{split}
B_3 &=- \frac{-3}{16 \pi ^2}\Bigl( \left\langle 1, 1 \middle\vert R^{1}\middle\vert 1,0 \right\rangle ^*\left\langle 0 ,1 \middle\vert R^{1}\middle\vert 1,0 \right\rangle + \left\langle 0, 1 \middle\vert R^{1}\middle\vert 1,0 \right\rangle ^*\left\langle 1 ,1 \middle\vert R^{1}\middle\vert 1,0 \right\rangle
   \Bigr).
 \end{split}
\end{equation}

\section{Description of ($n-\gamma $) reactions on nuclei with  spin $I=3/2$}
\label{App32}

For nuclei with spin $3/2$ , and with unknown spin of p-wave resonance, we have to consider two cases for the resonance with spin 1 and 2. Taking in mind $^{132}$Xe as a target, for  spin 1  the correlation coefficients for $M1$ ($g=1$ and $p=0$) and $E1$ ($g=1$ and $p=1$) $\gamma$ transitions correspond to the decay to the ground state of   $0^+$, while for  spin 2 the decay is going to the first exited state ($667.5$  keV) of $^{132}$Xe  $2^+$ \cite{XeNgamma1971,XeNgamma1988,XeNgamma2005} .

Thus for the p-wave resonance spin 1 we obtain:

\begin{eqnarray}\label{A0320}
\nonumber
 A^{J=1}_0&=& \frac{1}{16 \pi ^2}\left( 3\middle\vert \left\langle 0, 1 \middle\vert R^{1}\middle\vert 1,0 \right\rangle  \middle\vert^2+3 |\left\langle 1, 1 \middle\vert R^{1}\middle\vert 1,1 \right\rangle |^2 \right. \\
 &+& \left. 5\middle\vert \left\langle 0, 1 \middle\vert R^{2}\middle\vert 2,0 \right\rangle  \middle\vert^2+5 |\left\langle 1, 1 \middle\vert R^{2}\middle\vert 2,1 \right\rangle |^2  \right),
\end{eqnarray}

\begin{equation}\label{A1320}
\begin{split}
A^{J=1}_1&=-\frac{3\sqrt{3}}{16 \sqrt{2} \pi ^2} \Bigl([ \left\langle 1, 1 \middle\vert R^{1}\middle\vert 1,1 \right\rangle ^*\left\langle 0 , 1 \middle\vert R^{1}\middle\vert 1,0 \right\rangle +\left\langle 0, 1 \middle\vert R^{1}\middle\vert 1,0 \right\rangle ^* \left\langle 1, 1 \middle\vert R^{1}\middle\vert 1,1 \right\rangle]  \\
&+[ \left\langle 0, 1 \middle\vert R^{1}\middle\vert 1,1 \right\rangle ^*\left\langle 1 ,1 \middle\vert R^{1}\middle\vert 1,0 \right\rangle +\left\langle 1, 1 \middle\vert R^{1}\middle\vert 1,0 \right\rangle ^*\left\langle 0 ,1 \middle\vert R^{1}\middle\vert 1,1 \right\rangle ]\Bigr) ,
 \end{split}
\end{equation}

\begin{equation}\label{A2320}
\begin{split}
A^{J=1}_2 &=\frac{9}{320 \pi ^2} \Bigl(5 |\left\langle 0, 1 \middle\vert R^{1}\middle\vert 1,1 \right\rangle |^2 + 5 |\left\langle 1, 1 \middle\vert R^{1}\middle\vert 1,1 \right\rangle |^2 \\
&- |\left\langle 0, 1 \middle\vert R^{1}\middle\vert 2,1 \right\rangle |^2 -  |\left\langle 1, 1 \middle\vert R^{1}\middle\vert 2,1 \right\rangle |^2
 \Bigr) ,
 \end{split}
\end{equation}

\begin{equation}\label{B1320}
\begin{split}
B^{J=1}_1 & = \frac{3i}{32 \pi ^2}\sqrt{\frac{3}{2}}  \Bigl( \sqrt{5} \bigl( \left\langle 1, 1 \middle\vert R^{1}\middle\vert 2,1 \right\rangle ^*\left\langle 0,1 \middle\vert R^{1}\middle\vert 1,0 \right\rangle- \left\langle 0, 1 \middle\vert R^{1}\middle\vert 1,0 \right\rangle ^*\left\langle 1 , 1 \middle\vert R^{1}\middle\vert 2,1 \right\rangle \bigr)\\
&+  \bigl(\left\langle 1, 1 \middle\vert R^{1}\middle\vert 1,1 \right\rangle ^*\left\langle 0,1 \middle\vert R^{1}\middle\vert 1,0 \right\rangle -\left\langle 0, 1 \middle\vert R^{1}\middle\vert 1,0 \right\rangle ^*\left\langle 1,1 \middle\vert R^{1}\middle\vert 1,1 \right\rangle \bigr)\\
 &+ \bigl( \left\langle 0, 1 \middle\vert R^{1}\middle\vert 1,1 \right\rangle ^*\left\langle 1 ,1 \middle\vert R^{1}\middle\vert 1,0 \right\rangle -\left\langle 1, 1 \middle\vert R^{1}\middle\vert 1,0 \right\rangle ^*\left\langle 0 ,1 \middle\vert R^{1}\middle\vert 1,1 \right\rangle \bigr) \\
 &+\sqrt{5}\bigl(\left\langle 0, 1 \middle\vert R^{1}\middle\vert 2,1 \right\rangle ^*\left\langle 1 ,1 \middle\vert R^{1}\middle\vert 1,0 \right\rangle -\left\langle 1, 1 \middle\vert R^{1}\middle\vert 1,0 \right\rangle ^*\left\langle 0 ,1 \middle\vert R^{1}\middle\vert 2,1 \right\rangle  \bigr)  \Bigr) ,
 \end{split}
\end{equation}

\begin{equation}\label{B2320}
\begin{split}
B^{J=1}_2 &= \frac{1}{16 \sqrt{2} \pi ^2} \Bigl(\bigl(\sqrt{15}\left\langle 0 , 1 \middle\vert R^{1}\middle\vert 2,1 \right\rangle ^*\left\langle 0,1 \middle\vert R^{1}\middle\vert 1,0 \right\rangle +\left\langle 0,1 \middle\vert R^{1}\middle\vert 1,0 \right\rangle ^*\left\langle 0 , 1 \middle\vert R^{1}\middle\vert 2,1 \right\rangle \bigr) \\
&-\sqrt{3}\bigl(\left\langle 0 , 1 \middle\vert R^{1}\middle\vert 1,1 \right\rangle ^*\left\langle 0,1 \middle\vert R^{1}\middle\vert 1,0 \right\rangle +\left\langle 0,1 \middle\vert R^{1}\middle\vert 1,0 \right\rangle ^*\left\langle 0 , 1 \middle\vert R^{1}\middle\vert 1,1 \right\rangle \bigr) \\
 &-5\bigl( \left\langle 0, 1 \middle\vert R^{2}\middle\vert 2,0 \right\rangle ^*\left\langle 0 , 1 \middle\vert R^{2}\middle\vert 1,1 \right\rangle +\left\langle 0 , 1 \middle\vert R^{2}\middle\vert 1,1 \right\rangle ^*\left\langle 0, 1 \middle\vert R^{2}\middle\vert 2,0 \right\rangle \bigr) \\
 &+5\bigl( \left\langle 0, 1 \middle\vert R^{2}\middle\vert 2,1 \right\rangle ^*\left\langle 0 , 1 \middle\vert R^{2}\middle\vert 2,0 \right\rangle +\left\langle 0 , 1 \middle\vert R^{2}\middle\vert 2,0 \right\rangle ^*\left\langle 0, 1 \middle\vert R^{2}\middle\vert 2,1 \right\rangle \bigr) \\
 &-\sqrt{3}  \bigl( \left\langle 1 ,1 \middle\vert R^{1}\middle\vert 1,1 \right\rangle ^*\left\langle 1, 1 \middle\vert R^{1}\middle\vert 1,0 \right\rangle +\left\langle 1, 1 \middle\vert R^{1}\middle\vert 1,0 \right\rangle ^*\left\langle 1 ,1 \middle\vert R^{1}\middle\vert 1,1 \right\rangle \bigr) \\
  &+\sqrt{15}  \bigl( \left\langle 1 ,1 \middle\vert R^{1}\middle\vert 2,1 \right\rangle ^*\left\langle 1, 1 \middle\vert R^{1}\middle\vert 1,0 \right\rangle +\left\langle 1, 1 \middle\vert R^{1}\middle\vert 1,0 \right\rangle ^*\left\langle 1 ,1 \middle\vert R^{1}\middle\vert 2,1 \right\rangle \bigr) \\
    &-5  \bigl( \left\langle 1 ,1 \middle\vert R^{2}\middle\vert 2,0 \right\rangle ^*\left\langle 1, 1 \middle\vert R^{2}\middle\vert 1,1 \right\rangle +\left\langle 1, 1 \middle\vert R^{2}\middle\vert 1,1 \right\rangle ^*\left\langle 1 ,1 \middle\vert R^{2}\middle\vert 2,0 \right\rangle \bigr) \\
           &+ 5 \bigl( \left\langle 1 ,1 \middle\vert R^{2}\middle\vert 2,1 \right\rangle ^*\left\langle 1 ,1 \middle\vert R^{2}\middle\vert 2,0 \right\rangle +\left\langle 1 ,1 \middle\vert R^{2}\middle\vert 2,0 \right\rangle ^*\left\langle 1 ,1 \middle\vert R^{2}\middle\vert 2,1 \right\rangle \bigr) \Bigr),
 \end{split}
\end{equation}

\begin{equation}\label{B3320}
\begin{split}
B^{J=1}_3 &= -\frac{3}{32 \pi ^2}\Bigl( \left\langle 1, 1 \middle\vert R^{1}\middle\vert 1,0 \right\rangle ^*\left\langle 0 ,1 \middle\vert R^{1}\middle\vert 1,0 \right\rangle + \left\langle 0, 1 \middle\vert R^{1}\middle\vert 1,0 \right\rangle ^*\left\langle 1 ,1 \middle\vert R^{1}\middle\vert 1,0 \right\rangle
   \Bigr).
 \end{split}
\end{equation}

For case of the p-wave resonance with spin 2 we obtain:

\begin{eqnarray}\label{A0322}
\nonumber
 A^{J=2}_0&=& \frac{1}{16 \pi ^2}\left( 3\middle\vert \left\langle 0, 1 \middle\vert R^{1}\middle\vert 1,0 \right\rangle  \middle\vert^2+3 |\left\langle 1, 1 \middle\vert R^{1}\middle\vert 1,1 \right\rangle |^2 \right. \\
 &+& \left. 5\middle\vert \left\langle 0, 1 \middle\vert R^{2}\middle\vert 2,0 \right\rangle  \middle\vert^2+5 |\left\langle 1, 1 \middle\vert R^{2}\middle\vert 2,1 \right\rangle |^2  \right),
\end{eqnarray}

\begin{equation}\label{A1322}
\begin{split}
A^{J=2}_1&=-\frac{1}{32 \sqrt{2} \pi ^2} \Bigl(-3\sqrt{3}[ \left\langle 1, 1 \middle\vert R^{1}\middle\vert 1,1 \right\rangle ^*\left\langle 0 , 1 \middle\vert R^{1}\middle\vert 1,0 \right\rangle +\left\langle 0, 1 \middle\vert R^{1}\middle\vert 1,0 \right\rangle ^* \left\langle 1, 1 \middle\vert R^{1}\middle\vert 1,1 \right\rangle]  \\
&-3\sqrt{3}[ \left\langle 0, 1 \middle\vert R^{1}\middle\vert 1,1 \right\rangle ^*\left\langle 1 ,1 \middle\vert R^{1}\middle\vert 1,0 \right\rangle +\left\langle 1, 1 \middle\vert R^{1}\middle\vert 1,0 \right\rangle ^*\left\langle 0 ,1 \middle\vert R^{1}\middle\vert 1,1 \right\rangle ] \\
&-3\sqrt{15}[ \left\langle 1, 1 \middle\vert R^{2}\middle\vert 1,1 \right\rangle ^*\left\langle 0 ,1 \middle\vert R^{1}\middle\vert 1,0 \right\rangle +\left\langle 0, 1 \middle\vert R^{1}\middle\vert 1,0 \right\rangle ^*\left\langle 1 ,1 \middle\vert R^{2}\middle\vert 1,1 \right\rangle ] \\
&-3\sqrt{15}[ \left\langle 1, 1 \middle\vert R^{1}\middle\vert 1,0 \right\rangle ^*\left\langle 0 ,1 \middle\vert R^{2}\middle\vert 1,1 \right\rangle +\left\langle 0, 1 \middle\vert R^{2}\middle\vert 1,1 \right\rangle ^*\left\langle 1 ,1 \middle\vert R^{1}\middle\vert 1,0 \right\rangle ] \\
&+9[ \left\langle 1, 1 \middle\vert R^{2}\middle\vert 2,0 \right\rangle ^*\left\langle 0 ,1 \middle\vert R^{1}\middle\vert 2,1 \right\rangle +\left\langle 0, 1 \middle\vert R^{1}\middle\vert 2,1 \right\rangle ^*\left\langle 1 ,1 \middle\vert R^{2}\middle\vert 2,0 \right\rangle ] \\
&+9[ \left\langle 1, 1 \middle\vert R^{1}\middle\vert 2,1 \right\rangle ^*\left\langle 0 ,1 \middle\vert R^{2}\middle\vert 2,0 \right\rangle +\left\langle 0, 1 \middle\vert R^{2}\middle\vert 2,0 \right\rangle ^*\left\langle 1 ,1 \middle\vert R^{1}\middle\vert 2,1 \right\rangle ] \\
&+5[ \left\langle 1, 1 \middle\vert R^{2}\middle\vert 2,1 \right\rangle ^*\left\langle 0 ,1 \middle\vert R^{2}\middle\vert 2,0 \right\rangle +\left\langle 0, 1 \middle\vert R^{2}\middle\vert 2,0 \right\rangle ^*\left\langle 1 ,1 \middle\vert R^{2}\middle\vert 2,1 \right\rangle ] \\
&+5[ \left\langle 1, 1 \middle\vert R^{2}\middle\vert 2,0 \right\rangle ^*\left\langle 0 ,1 \middle\vert R^{2}\middle\vert 2,1 \right\rangle +\left\langle 0, 1 \middle\vert R^{2}\middle\vert 2,1 \right\rangle ^*\left\langle 1 ,1 \middle\vert R^{2}\middle\vert 2,0 \right\rangle ]
\Bigr),
\end{split}
\end{equation}

\begin{equation}\label{A2322}
\begin{split}
A^{J=2}_2 &=\frac{3}{3200 \pi ^2} \Bigl(15 |\left\langle 0, 1 \middle\vert R^{1}\middle\vert 1,1 \right\rangle |^2 + 15 |\left\langle 1, 1 \middle\vert R^{1}\middle\vert 1,1 \right\rangle |^2 \\
&- 3|\left\langle 0, 1 \middle\vert R^{1}\middle\vert 2,1 \right\rangle |^2 - 3 |\left\langle 1, 1 \middle\vert R^{1}\middle\vert 2,1 \right\rangle |^2 \\
&+175|\left\langle 0, 1 \middle\vert R^{2}\middle\vert 1,1 \right\rangle |^2 +175 |\left\langle 1, 1 \middle\vert R^{2}\middle\vert 1,1 \right\rangle |^2 \\
&-175|\left\langle 0, 1 \middle\vert R^{2}\middle\vert 2,1 \right\rangle |^2 -175 |\left\langle 1, 1 \middle\vert R^{2}\middle\vert 2,1 \right\rangle |^2 \\
&+45\sqrt{5}[ \left\langle 0, 1 \middle\vert R^{2}\middle\vert 1,1 \right\rangle ^*\left\langle 0 ,1 \middle\vert R^{1}\middle\vert 1,1 \right\rangle +\left\langle 0, 1 \middle\vert R^{1}\middle\vert 1,1 \right\rangle ^*\left\langle 0 ,1 \middle\vert R^{2}\middle\vert 1,1 \right\rangle ] \\
&-45[ \left\langle 0, 1 \middle\vert R^{2}\middle\vert 2,1 \right\rangle ^*\left\langle 0 ,1 \middle\vert R^{1}\middle\vert 2,1 \right\rangle +\left\langle 0, 1 \middle\vert R^{1}\middle\vert 2,1 \right\rangle ^*\left\langle 0 ,1 \middle\vert R^{2}\middle\vert 2,1 \right\rangle ] \\
&+45\sqrt{5}[ \left\langle 1, 1 \middle\vert R^{2}\middle\vert 1,1 \right\rangle ^*\left\langle 1 ,1 \middle\vert R^{1}\middle\vert 1,1 \right\rangle +\left\langle 1, 1 \middle\vert R^{1}\middle\vert 1,1 \right\rangle ^*\left\langle 1 ,1 \middle\vert R^{2}\middle\vert 1,1 \right\rangle ] \\
&-45[ \left\langle 1, 1 \middle\vert R^{2}\middle\vert 2,1 \right\rangle ^*\left\langle 1 ,1 \middle\vert R^{1}\middle\vert 2,1 \right\rangle +\left\langle 1, 1 \middle\vert R^{1}\middle\vert 2,1 \right\rangle ^*\left\langle 1 ,1 \middle\vert R^{2}\middle\vert 2,1 \right\rangle ]
 \Bigr) ,
 \end{split}
\end{equation}

\begin{equation}\label{B1322}
\begin{split}
B^{J=2}_1 & = -\frac{i}{64 \sqrt{2} \pi ^2}  \Bigl( 3\sqrt{15} \bigl( \left\langle 1, 1 \middle\vert R^{1}\middle\vert 2,1 \right\rangle ^*\left\langle 0,1 \middle\vert R^{1}\middle\vert 1,0 \right\rangle- \left\langle 0, 1 \middle\vert R^{1}\middle\vert 1,0 \right\rangle ^*\left\langle 1 , 1 \middle\vert R^{1}\middle\vert 2,1 \right\rangle \bigr)\\
&+3\sqrt{3}  \bigl(\left\langle 1, 1 \middle\vert R^{1}\middle\vert 1,1 \right\rangle ^*\left\langle 0,1 \middle\vert R^{1}\middle\vert 1,0 \right\rangle -\left\langle 0, 1 \middle\vert R^{1}\middle\vert 1,0 \right\rangle ^*\left\langle 1,1 \middle\vert R^{1}\middle\vert 1,1 \right\rangle \bigr)\\
 &+3\sqrt{3} \bigl( \left\langle 0, 1 \middle\vert R^{1}\middle\vert 1,1 \right\rangle ^*\left\langle 1 ,1 \middle\vert R^{1}\middle\vert 1,0 \right\rangle -\left\langle 1, 1 \middle\vert R^{1}\middle\vert 1,0 \right\rangle ^*\left\langle 0 ,1 \middle\vert R^{1}\middle\vert 1,1 \right\rangle \bigr) \\
  &+3\sqrt{15}\bigl(\left\langle 0, 1 \middle\vert R^{1}\middle\vert 2,1 \right\rangle ^*\left\langle 1 ,1 \middle\vert R^{1}\middle\vert 1,0 \right\rangle -\left\langle 1, 1 \middle\vert R^{1}\middle\vert 1,0 \right\rangle ^*\left\langle 0 ,1 \middle\vert R^{1}\middle\vert 2,1 \right\rangle  \bigr) \\
  &+3\sqrt{15}\bigl(\left\langle 0, 1 \middle\vert R^{1}\middle\vert 1,0 \right\rangle ^*\left\langle 1 ,1 \middle\vert R^{1}\middle\vert 1,1 \right\rangle -\left\langle 1, 1 \middle\vert R^{2}\middle\vert 1,1 \right\rangle ^*\left\langle 0 ,1 \middle\vert R^{1}\middle\vert 1,0 \right\rangle  \bigr) \\
  &+9\sqrt{15}\bigl(\left\langle 1, 1 \middle\vert R^{2}\middle\vert 2,1 \right\rangle ^*\left\langle 0 ,1 \middle\vert R^{1}\middle\vert 1,0 \right\rangle -\left\langle 0, 1 \middle\vert R^{1}\middle\vert 1,0 \right\rangle ^*\left\langle 1 ,1 \middle\vert R^{2}\middle\vert 2,1 \right\rangle  \bigr) \\
  &+9\sqrt{5}\bigl(\left\langle 1, 1 \middle\vert R^{2}\middle\vert 2,0 \right\rangle ^*\left\langle 0 ,1 \middle\vert R^{1}\middle\vert 1,1 \right\rangle -\left\langle 0, 1 \middle\vert R^{1}\middle\vert 1,1 \right\rangle ^*\left\langle 1 ,1 \middle\vert R^{2}\middle\vert 2,0 \right\rangle  \bigr) \\
  &+27\bigl(\left\langle 1, 1 \middle\vert R^{2}\middle\vert 2,0 \right\rangle ^*\left\langle 0 ,1 \middle\vert R^{1}\middle\vert 2,1 \right\rangle -\left\langle 0, 1 \middle\vert R^{1}\middle\vert 2,1 \right\rangle ^*\left\langle 1 ,1 \middle\vert R^{2}\middle\vert 2,0 \right\rangle  \bigr) \\
    &+3\sqrt{15}\bigl(\left\langle 1, 1 \middle\vert R^{1}\middle\vert 1,0 \right\rangle ^*\left\langle 0 ,1 \middle\vert R^{2}\middle\vert 1,1 \right\rangle -\left\langle 0, 1 \middle\vert R^{2}\middle\vert 1,1 \right\rangle ^*\left\langle 1 ,1 \middle\vert R^{1}\middle\vert 1,0 \right\rangle  \bigr) \\
    &+15\bigl(\left\langle 1, 1 \middle\vert R^{2}\middle\vert 2,0 \right\rangle ^*\left\langle 0 ,1 \middle\vert R^{2}\middle\vert 1,1 \right\rangle -\left\langle 0, 1 \middle\vert R^{2}\middle\vert 1,1 \right\rangle ^*\left\langle 1 ,1 \middle\vert R^{2}\middle\vert 2,0 \right\rangle  \bigr) \\
    &+9\sqrt{5}\bigl(\left\langle 0, 1 \middle\vert R^{2}\middle\vert 2,0 \right\rangle ^*\left\langle 1 ,1 \middle\vert R^{1}\middle\vert 1,1 \right\rangle -\left\langle 1, 1 \middle\vert R^{1}\middle\vert 1,1 \right\rangle ^*\left\langle 0 ,1 \middle\vert R^{2}\middle\vert 2,0 \right\rangle  \bigr) \\
    &+27\bigl(\left\langle 1, 1 \middle\vert R^{1}\middle\vert 2,1 \right\rangle ^*\left\langle 0 ,1 \middle\vert R^{2}\middle\vert 2,0 \right\rangle -\left\langle 0, 1 \middle\vert R^{2}\middle\vert 2,0 \right\rangle ^*\left\langle 1 ,1 \middle\vert R^{1}\middle\vert 2,1 \right\rangle  \bigr) \\
    &+15\bigl(\left\langle 0, 1 \middle\vert R^{2}\middle\vert 2,0 \right\rangle ^*\left\langle 1 ,1 \middle\vert R^{2}\middle\vert 1,1 \right\rangle -\left\langle 1, 1 \middle\vert R^{2}\middle\vert 1,1 \right\rangle ^*\left\langle 0 ,1 \middle\vert R^{2}\middle\vert 2,0 \right\rangle  \bigr) \\
    &+9\sqrt{15}\bigl(\left\langle 0, 1 \middle\vert R^{2}\middle\vert 2,1 \right\rangle ^*\left\langle 1 ,1 \middle\vert R^{1}\middle\vert 1,0 \right\rangle -\left\langle 1, 1 \middle\vert R^{1}\middle\vert 1,0 \right\rangle ^*\left\langle 0 ,1 \middle\vert R^{2}\middle\vert 2,1 \right\rangle  \bigr) \\
    &+5\bigl(\left\langle 1, 1 \middle\vert R^{2}\middle\vert 2,1 \right\rangle ^*\left\langle 0 ,1 \middle\vert R^{2}\middle\vert 2,0 \right\rangle -\left\langle 0, 1 \middle\vert R^{2}\middle\vert 2,0 \right\rangle ^*\left\langle 1 ,1 \middle\vert R^{2}\middle\vert 2,1 \right\rangle  \bigr) \\
    &+5\bigl(\left\langle 0, 1 \middle\vert R^{2}\middle\vert 2,1 \right\rangle ^*\left\langle 1 ,1 \middle\vert R^{2}\middle\vert 2,0 \right\rangle -\left\langle 1, 1 \middle\vert R^{2}\middle\vert 2,0 \right\rangle ^*\left\langle 0 ,1 \middle\vert R^{2}\middle\vert 2,1 \right\rangle  \bigr)
  \Bigr) ,
 \end{split}
\end{equation}

\begin{equation}\label{B2322}
\begin{split}
B^{J=2}_2 &= \frac{1}{16 \sqrt{2} \pi ^2} \Bigl(\bigl(\sqrt{15}\left\langle 0 , 1 \middle\vert R^{1}\middle\vert 2,1 \right\rangle ^*\left\langle 0,1 \middle\vert R^{1}\middle\vert 1,0 \right\rangle +\left\langle 0,1 \middle\vert R^{1}\middle\vert 1,0 \right\rangle ^*\left\langle 0 , 1 \middle\vert R^{1}\middle\vert 2,1 \right\rangle \bigr) \\
&-\sqrt{3}\bigl(\left\langle 0 , 1 \middle\vert R^{1}\middle\vert 1,1 \right\rangle ^*\left\langle 0,1 \middle\vert R^{1}\middle\vert 1,0 \right\rangle +\left\langle 0,1 \middle\vert R^{1}\middle\vert 1,0 \right\rangle ^*\left\langle 0 , 1 \middle\vert R^{1}\middle\vert 1,1 \right\rangle \bigr) \\
 &-5\bigl( \left\langle 0, 1 \middle\vert R^{2}\middle\vert 2,0 \right\rangle ^*\left\langle 0 , 1 \middle\vert R^{2}\middle\vert 1,1 \right\rangle +\left\langle 0 , 1 \middle\vert R^{2}\middle\vert 1,1 \right\rangle ^*\left\langle 0, 1 \middle\vert R^{2}\middle\vert 2,0 \right\rangle \bigr) \\
 &+5\bigl( \left\langle 0, 1 \middle\vert R^{2}\middle\vert 2,1 \right\rangle ^*\left\langle 0 , 1 \middle\vert R^{2}\middle\vert 2,0 \right\rangle +\left\langle 0 , 1 \middle\vert R^{2}\middle\vert 2,0 \right\rangle ^*\left\langle 0, 1 \middle\vert R^{2}\middle\vert 2,1 \right\rangle \bigr) \\
 &-\sqrt{3}  \bigl( \left\langle 1 ,1 \middle\vert R^{1}\middle\vert 1,1 \right\rangle ^*\left\langle 1, 1 \middle\vert R^{1}\middle\vert 1,0 \right\rangle +\left\langle 1, 1 \middle\vert R^{1}\middle\vert 1,0 \right\rangle ^*\left\langle 1 ,1 \middle\vert R^{1}\middle\vert 1,1 \right\rangle \bigr) \\
  &+\sqrt{15}  \bigl( \left\langle 1 ,1 \middle\vert R^{1}\middle\vert 2,1 \right\rangle ^*\left\langle 1, 1 \middle\vert R^{1}\middle\vert 1,0 \right\rangle +\left\langle 1, 1 \middle\vert R^{1}\middle\vert 1,0 \right\rangle ^*\left\langle 1 ,1 \middle\vert R^{1}\middle\vert 2,1 \right\rangle \bigr) \\
    &-5  \bigl( \left\langle 1 ,1 \middle\vert R^{2}\middle\vert 2,0 \right\rangle ^*\left\langle 1, 1 \middle\vert R^{2}\middle\vert 1,1 \right\rangle +\left\langle 1, 1 \middle\vert R^{2}\middle\vert 1,1 \right\rangle ^*\left\langle 1 ,1 \middle\vert R^{2}\middle\vert 2,0 \right\rangle \bigr) \\
           &+ 5 \bigl( \left\langle 1 ,1 \middle\vert R^{2}\middle\vert 2,1 \right\rangle ^*\left\langle 1 ,1 \middle\vert R^{2}\middle\vert 2,0 \right\rangle +\left\langle 1 ,1 \middle\vert R^{2}\middle\vert 2,0 \right\rangle ^*\left\langle 1 ,1 \middle\vert R^{2}\middle\vert 2,1 \right\rangle \bigr) \Bigr),
 \end{split}
\end{equation}

\begin{equation}\label{B3322}
\begin{split}
B^{J=2}_3 &= \frac{1}{64 \pi ^2}\Bigl(3 \bigl( \left\langle 1, 1 \middle\vert R^{1}\middle\vert 1,0 \right\rangle ^*\left\langle 0 ,1 \middle\vert R^{1}\middle\vert 1,0 \right\rangle + \left\langle 0, 1 \middle\vert R^{1}\middle\vert 1,0 \right\rangle ^*\left\langle 1 ,1 \middle\vert R^{1}\middle\vert 1,0 \right\rangle \bigr) \\
&+3 \sqrt{15}\bigl( \left\langle 1, 1 \middle\vert R^{2}\middle\vert 2,0 \right\rangle ^*\left\langle 0 ,1 \middle\vert R^{1}\middle\vert 1,0 \right\rangle + \left\langle 0, 1 \middle\vert R^{1}\middle\vert 1,0 \right\rangle ^*\left\langle 1 ,1 \middle\vert R^{2}\middle\vert 2,0 \right\rangle  \bigr)\\
&+3 \sqrt{15}\bigl( \left\langle 1, 1 \middle\vert R^{1}\middle\vert 1,0 \right\rangle ^*\left\langle 0 ,1 \middle\vert R^{2}\middle\vert 2,0 \right\rangle + \left\langle 0, 1 \middle\vert R^{2}\middle\vert 2,0 \right\rangle ^*\left\langle 1 ,1 \middle\vert R^{1}\middle\vert 1,0 \right\rangle \bigr) \\
&+5\bigl( \left\langle 1, 1 \middle\vert R^{2}\middle\vert 2,0 \right\rangle ^*\left\langle 0 ,1 \middle\vert R^{2}\middle\vert 2,0 \right\rangle + \left\langle 0, 1 \middle\vert R^{2}\middle\vert 2,0 \right\rangle ^*\left\langle 1 ,1 \middle\vert R^{2}\middle\vert 2,0 \right\rangle \bigr)
   \Bigr).
 \end{split}
\end{equation}

\section{Description of ($n-\gamma $) reactions on nuclei with  spin $I=7/2$}
\label{App72}

The measured
correlation coefficients for a target with spin $7/2$ for $M1$ ($g=1$ and $p=0$) and $E1$ ($g=1$ and $p=1$) $\gamma$ transitions to the ground state with spin 3 can be written as

\begin{eqnarray}\label{A072}
\nonumber
 A_0&=& \frac{1}{16 \pi ^2}\left(7 \middle\vert \left\langle 1, 1 \middle\vert R^{3}\middle\vert 3,0 \right\rangle  \middle\vert^2+9 |\left\langle 1, 1 \middle\vert R^{4}\middle\vert 4,0 \right\rangle |^2 \right. \\
 &+& \left.7 \middle\vert \left\langle 0, 1 \middle\vert R^{3}\middle\vert 3,1 \right\rangle  \middle\vert^2+9 |\left\langle 0, 1 \middle\vert R^{4}\middle\vert 4,1 \right\rangle |^2  \right),
\end{eqnarray}

\begin{equation}\label{A172}
\begin{split}
A_1&=-\frac{1}{64 \pi ^2} \Bigl(7[ \left\langle 1, 1 \middle\vert R^{3}\middle\vert 3,1 \right\rangle ^*\left\langle 0 , 1 \middle\vert R^{3}\middle\vert 3,0 \right\rangle +\left\langle 0, 1 \middle\vert R^{3}\middle\vert 3,0 \right\rangle ^* \left\langle 1, 1 \middle\vert R^{3}\middle\vert 3,1 \right\rangle]  \\
&+7[ \left\langle 1, 1 \middle\vert R^{3}\middle\vert 3,0 \right\rangle ^*\left\langle 0 , 1 \middle\vert R^{3}\middle\vert 3,1 \right\rangle +\left\langle 0, 1 \middle\vert R^{3}\middle\vert 3,1 \right\rangle ^* \left\langle 1, 1 \middle\vert R^{3}\middle\vert 3,0 \right\rangle]  \\
&+9  \sqrt{7}[ \left\langle 1, 1 \middle\vert R^{3}\middle\vert 4,1 \right\rangle ^*\left\langle 0 ,1 \middle\vert R^{4}\middle\vert 4,0 \right\rangle + \left\langle 0, 1 \middle\vert R^{4}\middle\vert 4,0 \right\rangle ^*\left\langle 1, 1 \middle\vert R^{3}\middle\vert 4,1 \right\rangle ] \\
&+9  \sqrt{7}[ \left\langle 0, 1 \middle\vert R^{3}\middle\vert 4,1 \right\rangle ^*\left\langle 1 ,1 \middle\vert R^{4}\middle\vert 4,0 \right\rangle + \left\langle 1, 1 \middle\vert R^{4}\middle\vert 4,0 \right\rangle ^*\left\langle 0, 1 \middle\vert R^{3}\middle\vert 4,1 \right\rangle ] \\
&-27 [ \left\langle 1, 1 \middle\vert R^{4}\middle\vert 3,1 \right\rangle ^*\left\langle 0 , 1 \middle\vert R^{3}\middle\vert 3,0 \right\rangle+\left\langle 0, 1 \middle\vert R^{3}\middle\vert 3,0 \right\rangle ^*\left\langle 1, 1 \middle\vert R^{4}\middle\vert 3,1 \right\rangle ] \\
&-27 [ \left\langle 0, 1 \middle\vert R^{4}\middle\vert 3,1 \right\rangle ^*\left\langle 1 , 1 \middle\vert R^{3}\middle\vert 3,0 \right\rangle+\left\langle 1, 1 \middle\vert R^{3}\middle\vert 3,0 \right\rangle ^*\left\langle 0, 1 \middle\vert R^{4}\middle\vert 3,1 \right\rangle ] \\
&+9 \sqrt{15}[ \left\langle 1, 1 \middle\vert R^{4}\middle\vert 4,1 \right\rangle ^*\left\langle 0 ,1 \middle\vert R^{4}\middle\vert 4,0 \right\rangle +\left\langle 0, 1 \middle\vert R^{4}\middle\vert 4,0 \right\rangle ^*\left\langle 1 ,1 \middle\vert R^{4}\middle\vert 4,1 \right\rangle ] \\
&+9 \sqrt{15}[ \left\langle 0, 1 \middle\vert R^{4}\middle\vert 4,1 \right\rangle ^*\left\langle 1 ,1 \middle\vert R^{4}\middle\vert 4,0 \right\rangle +\left\langle 1, 1 \middle\vert R^{4}\middle\vert 4,0 \right\rangle ^*\left\langle 0 ,1 \middle\vert R^{4}\middle\vert 4,1 \right\rangle ]\Bigr) ,
 \end{split}
\end{equation}

\begin{equation}\label{A272}
\begin{split}
A_2 &=\frac{3}{256 \pi ^2} \Bigl(7\bigl( \left\langle 0, 1 \middle\vert R^{3}\middle\vert 4,1 \right\rangle ^*\left\langle 0, 1 \middle\vert R^{3}\middle\vert 4,1 \right\rangle + \left\langle 1, 1 \middle\vert R^{3}\middle\vert 4,1 \right\rangle ^*\left\langle 1, 1 \middle\vert R^{3}\middle\vert 4,1 \right\rangle \bigr) \\
&-21\bigl(\left\langle 1,1 \middle\vert R^{3}\middle\vert 3,1 \right\rangle ^*\left\langle 1,1 \middle\vert R^{3}\middle\vert 3,1 \right\rangle -\left\langle 0,1 \middle\vert R^{3}\middle\vert 3,1 \right\rangle ^*\left\langle 0,1 \middle\vert R^{3}\middle\vert 3,1 \right\rangle \bigr) \\
&-27 \bigl( \left\langle 0, 1 \middle\vert R^{4}\middle\vert 3,1 \right\rangle ^*\left\langle 0,1 \middle\vert R^{3}\middle\vert 3,1 \right\rangle +\left\langle 0,1 \middle\vert R^{3}\middle\vert 3,1 \right\rangle ^*\left\langle 0, 1 \middle\vert R^{4}\middle\vert 3,1 \right\rangle \bigr)\\
&-27 \bigl( \left\langle 1, 1 \middle\vert R^{4}\middle\vert 3,1 \right\rangle ^*\left\langle 1,1 \middle\vert R^{3}\middle\vert 3,1 \right\rangle +\left\langle 1,1 \middle\vert R^{3}\middle\vert 3,1 \right\rangle ^*\left\langle , 1 \middle\vert R^{4}\middle\vert 3,1 \right\rangle \bigr)\\
&+9 \sqrt{\frac{21}{5}}  \bigl( \left\langle 0 ,1 \middle\vert R^{4}\middle\vert 4,1 \right\rangle ^*\left\langle 0, 1 \middle\vert R^{3}\middle\vert 4,1 \right\rangle + \left\langle 0, 1 \middle\vert R^{3}\middle\vert 4,1 \right\rangle ^*\left\langle 0 ,1 \middle\vert R^{4}\middle\vert 4,1 \right\rangle  \bigr) \\
&+9 \sqrt{\frac{21}{5}}  \bigl( \left\langle 1 ,1 \middle\vert R^{4}\middle\vert 4,1 \right\rangle ^*\left\langle 1, 1 \middle\vert R^{3}\middle\vert 4,1 \right\rangle + \left\langle 1, 1 \middle\vert R^{3}\middle\vert 4,1 \right\rangle ^*\left\langle 1 ,1 \middle\vert R^{4}\middle\vert 4,1 \right\rangle  \bigr) \\
&-\frac{99}{7}\bigl(  \left\langle 0, 1 \middle\vert R^{4}\middle\vert 3,1 \right\rangle ^*\left\langle 0, 1 \middle\vert R^{4}\middle\vert 3,1 \right\rangle +\left\langle 1, 1 \middle\vert R^{4}\middle\vert 3,1 \right\rangle ^*\left\langle 1, 1 \middle\vert R^{4}\middle\vert 3,1 \right\rangle \bigr) \\
&+\frac{99}{5}\bigl( \left\langle 0 ,1 \middle\vert R^{4}\middle\vert 4,1 \right\rangle ^*\left\langle 0 ,1 \middle\vert R^{4}\middle\vert 4,1 \right\rangle +\left\langle 1 ,1 \middle\vert R^{4}\middle\vert 4,1 \right\rangle ^*\left\langle 1 ,1 \middle\vert R^{4}\middle\vert 4,1 \right\rangle \bigr)
 \Bigr) ,
 \end{split}
\end{equation}

\begin{equation}\label{B172}
\begin{split}
B_1 & = -\frac{i}{256 \pi ^2}  \Bigl( 7 \bigl( \left\langle 0, 1 \middle\vert R^{3}\middle\vert 3,0 \right\rangle ^*\left\langle 1,1 \middle\vert R^{3}\middle\vert 3,1 \right\rangle- \left\langle 1, 1 \middle\vert R^{3}\middle\vert 3,1 \right\rangle ^*\left\langle 0 , 1 \middle\vert R^{3}\middle\vert 3,0 \right\rangle \bigr)\\
&+ 7 \bigl(\left\langle 0, 1 \middle\vert R^{3}\middle\vert 3,1 \right\rangle ^*\left\langle 1,1 \middle\vert R^{3}\middle\vert 3,0 \right\rangle -\left\langle 1, 1 \middle\vert R^{3}\middle\vert 3,0 \right\rangle ^*\left\langle 0,1 \middle\vert R^{3}\middle\vert 3,1 \right\rangle \bigr)\\
 &+21 \sqrt{3}\bigl(\left\langle 0, 1 \middle\vert R^{3}\middle\vert 3,0 \right\rangle ^*\left\langle 1 , 1 \middle\vert R^{3}\middle\vert 4,1 \right\rangle -\left\langle 1, 1 \middle\vert R^{3}\middle\vert 4,1 \right\rangle ^*\left\langle 0 , 1 \middle\vert R^{3}\middle\vert 3,0 \right\rangle  \bigr) \\
   &-21 \sqrt{3}\bigl( \left\langle 0, 1 \middle\vert R^{3}\middle\vert 4,1 \right\rangle ^*\left\langle 1 , 1 \middle\vert R^{3}\middle\vert 3,0 \right\rangle - \left\langle 1, 1 \middle\vert R^{3}\middle\vert 3,0 \right\rangle ^*\left\langle 0, 1 \middle\vert R^{3}\middle\vert 4,1 \right\rangle \bigr) \\
  &+81\bigl( \left\langle 0, 1 \middle\vert R^{3}\middle\vert 3,0 \right\rangle ^*\left\langle 1, 1 \middle\vert R^{4}\middle\vert 3,1 \right\rangle - \left\langle 1, 1 \middle\vert R^{4}\middle\vert 3,1 \right\rangle ^*\left\langle 0 , 1 \middle\vert R^{3}\middle\vert 3,0 \right\rangle \bigr) \\
   &-81\bigl(\left\langle 0, 1 \middle\vert R^{4}\middle\vert 3,1 \right\rangle ^*\left\langle 1, 1 \middle\vert R^{3}\middle\vert 3,0 \right\rangle - \left\langle 1, 1 \middle\vert R^{3}\middle\vert 3,0 \right\rangle ^*\left\langle 0, 1 \middle\vert R^{4}\middle\vert 3,1 \right\rangle  \\
&+27 \sqrt{35} \bigl( \left\langle 1, 1 \middle\vert R^{4}\middle\vert 4,1 \right\rangle ^*\left\langle 0 , 1 \middle\vert R^{3}\middle\vert 3,0 \right\rangle -\left\langle 0, 1 \middle\vert R^{3}\middle\vert 3,0 \right\rangle ^*\left\langle 1 ,1 \middle\vert R^{4}\middle\vert 4,1 \right\rangle \bigr) \\
&-27 \sqrt{35} \bigl(\left\langle 1, 1 \middle\vert R^{3}\middle\vert 3,0 \right\rangle ^*\left\langle 0 ,1 \middle\vert R^{4}\middle\vert 4,1 \right\rangle -\left\langle 0, 1 \middle\vert R^{4}\middle\vert 4,1 \right\rangle ^*\left\langle 1 ,1 \middle\vert R^{3}\middle\vert 3,0 \right\rangle \bigr) \\
&-27 \sqrt{21} \bigl( \left\langle 0, 1 \middle\vert R^{3}\middle\vert 3,1 \right\rangle ^*\left\langle 1,1 \middle\vert R^{4}\middle\vert 4,0 \right\rangle -\left\langle 1, 1 \middle\vert R^{4}\middle\vert 4,0 \right\rangle ^*\left\langle 0,1 \middle\vert R^{3}\middle\vert 3,1 \right\rangle  \bigr) \\
&+27 \sqrt{21} \bigl(\left\langle 0, 1 \middle\vert R^{4}\middle\vert 4,0 \right\rangle ^*\left\langle 1 ,1 \middle\vert R^{3}\middle\vert 3,1 \right\rangle -\left\langle 1, 1 \middle\vert R^{3}\middle\vert 3,1 \right\rangle ^*\left\langle 0 ,1 \middle\vert R^{4}\middle\vert 4,0 \right\rangle \bigr) \\
&-45 \sqrt{7}\bigl( \left\langle 1, 1 \middle\vert R^{4}\middle\vert 4,0 \right\rangle ^*\left\langle 0, 1 \middle\vert R^{3}\middle\vert 4,1 \right\rangle -\left\langle 0, 1 \middle\vert R^{3}\middle\vert 4,1 \right\rangle ^*\left\langle 1 ,1 \middle\vert R^{4}\middle\vert 4,0 \right\rangle \bigr) \\
&+45 \sqrt{7}\bigl( \left\langle 1, 1 \middle\vert R^{3}\middle\vert 4,1 \right\rangle ^*\left\langle 0 ,1 \middle\vert R^{4}\middle\vert 4,0 \right\rangle -\left\langle 0, 1 \middle\vert R^{4}\middle\vert 4,0 \right\rangle ^*\left\langle 1, 1 \middle\vert R^{3}\middle\vert 4,1 \right\rangle  \bigr) \\
&-45 \sqrt{21} \bigl(\left\langle 0, 1 \middle\vert R^{4}\middle\vert 3,1 \right\rangle ^*\left\langle 1 ,1 \middle\vert R^{4}\middle\vert 4,0 \right\rangle-\left\langle 1, 1 \middle\vert R^{4}\middle\vert 4,0 \right\rangle ^*\left\langle 0, 1 \middle\vert R^{4}\middle\vert 3,1 \right\rangle
 \bigr) \\
&+45 \sqrt{21} \bigl(\left\langle 0, 1 \middle\vert R^{4}\middle\vert 4,0 \right\rangle ^*\left\langle 1, 1 \middle\vert R^{4}\middle\vert 3,1 \right\rangle
-\left\langle 1, 1 \middle\vert R^{4}\middle\vert 3,1 \right\rangle ^*\left\langle 0 ,1 \middle\vert R^{4}\middle\vert 4,0 \right\rangle \bigr) \\
 &+9 \sqrt{15}\bigl( \left\langle 1, 1 \middle\vert R^{4}\middle\vert 4,1 \right\rangle ^*\left\langle 0 ,1 \middle\vert R^{4}\middle\vert 4,0 \right\rangle -\left\langle 0, 1 \middle\vert R^{4}\middle\vert 4,0 \right\rangle ^*\left\langle 1 ,1 \middle\vert R^{4}\middle\vert 4,1 \right\rangle \bigr) \\
 &-9 \sqrt{15}\bigl(\left\langle 1, 1 \middle\vert R^{4}\middle\vert 4,0 \right\rangle ^*\left\langle 0 ,1 \middle\vert R^{4}\middle\vert 4,1 \right\rangle -\left\langle 0, 1 \middle\vert R^{4}\middle\vert 4,1 \right\rangle ^*\left\langle 1 ,1 \middle\vert R^{4}\middle\vert 4,0 \right\rangle  \bigr)  \Bigr) ,
 \end{split}
\end{equation}

\begin{equation}\label{B272}
\begin{split}
B_2 &= \frac{1}{32 \pi ^2} \Bigl(- 7\bigl(\left\langle 0 , 1 \middle\vert R^{3}\middle\vert 3,0 \right\rangle ^*\left\langle 0,1 \middle\vert R^{3}\middle\vert 3,1 \right\rangle +\left\langle 0,1 \middle\vert R^{3}\middle\vert 3,1 \right\rangle ^*\left\langle 0 , 1 \middle\vert R^{3}\middle\vert 3,0 \right\rangle \bigr) \\
&-7\bigl(\left\langle 1 , 1 \middle\vert R^{3}\middle\vert 3,0 \right\rangle ^*\left\langle 1,1 \middle\vert R^{3}\middle\vert 3,1 \right\rangle +\left\langle 1,1 \middle\vert R^{3}\middle\vert 3,1 \right\rangle ^*\left\langle 1 , 1 \middle\vert R^{3}\middle\vert 3,0 \right\rangle \bigr) \\
 &+7 \sqrt{3}\bigl( \left\langle 0, 1 \middle\vert R^{3}\middle\vert 4,1 \right\rangle ^*\left\langle 0 , 1 \middle\vert R^{3}\middle\vert 3,0 \right\rangle +\left\langle 0 , 1 \middle\vert R^{3}\middle\vert 3,0 \right\rangle ^*\left\langle 0, 1 \middle\vert R^{3}\middle\vert 4,1 \right\rangle \bigr) \\
   &+7 \sqrt{3}\bigl( \left\langle 1, 1 \middle\vert R^{3}\middle\vert 4,1 \right\rangle ^*\left\langle 1 , 1 \middle\vert R^{3}\middle\vert 3,0 \right\rangle +\left\langle 1 , 1 \middle\vert R^{3}\middle\vert 3,0 \right\rangle ^*\left\langle 1, 1 \middle\vert R^{3}\middle\vert 4,1 \right\rangle \bigr) \\
   &-3 \sqrt{21} \bigl( \left\langle 0 ,1 \middle\vert R^{4}\middle\vert 4,0 \right\rangle ^*\left\langle 0, 1 \middle\vert R^{4}\middle\vert 3,1 \right\rangle +\left\langle 0, 1 \middle\vert R^{4}\middle\vert 3,1 \right\rangle ^*\left\langle 0 ,1 \middle\vert R^{4}\middle\vert 4,0 \right\rangle \bigr) \\
     &-3 \sqrt{21} \bigl( \left\langle 1 ,1 \middle\vert R^{4}\middle\vert 4,0 \right\rangle ^*\left\langle 1, 1 \middle\vert R^{4}\middle\vert 3,1 \right\rangle +\left\langle 1, 1 \middle\vert R^{4}\middle\vert 3,1 \right\rangle ^*\left\langle 1 ,1 \middle\vert R^{4}\middle\vert 4,0 \right\rangle \bigr) \\
      &+3 \sqrt{15} \bigl( \left\langle 0 ,1 \middle\vert R^{4}\middle\vert 4,0 \right\rangle ^*\left\langle 0 ,1 \middle\vert R^{4}\middle\vert 4,1 \right\rangle +\left\langle 0 ,1 \middle\vert R^{4}\middle\vert 4,1 \right\rangle ^*\left\langle 0 ,1 \middle\vert R^{4}\middle\vert 4,0 \right\rangle \bigr)\\
        &+3 \sqrt{15} \bigl( \left\langle 1 ,1 \middle\vert R^{4}\middle\vert 4,0 \right\rangle ^*\left\langle 1 ,1 \middle\vert R^{4}\middle\vert 4,1 \right\rangle +\left\langle 1 ,1 \middle\vert R^{4}\middle\vert 4,1 \right\rangle ^*\left\langle 1 ,1 \middle\vert R^{4}\middle\vert 4,0 \right\rangle \bigr) \Bigr),
 \end{split}
\end{equation}

\begin{equation}\label{B372}
\begin{split}
B_3 &= -\frac{1}{128 \pi ^2}\Bigl(-45\bigl( \left\langle 1, 1 \middle\vert R^{4}\middle\vert 4,0 \right\rangle ^*\left\langle 0 ,1 \middle\vert R^{4}\middle\vert 4,0 \right\rangle + \left\langle 0, 1 \middle\vert R^{4}\middle\vert 4,0 \right\rangle ^*\left\langle 1 ,1 \middle\vert R^{4}\middle\vert 4,0 \right\rangle \bigr)\\
 &+7\bigl(\left\langle 0, 1 \middle\vert R^{3}\middle\vert 3,0 \right\rangle ^*\left\langle 1 , 1 \middle\vert R^{3}\middle\vert 3,0 \right\rangle+\left\langle 1, 1 \middle\vert R^{3}\middle\vert 3,0 \right\rangle ^*\left\langle 0 , 1 \middle\vert R^{3}\middle\vert 3,0 \right\rangle \bigr) \\
  &-9 \sqrt{21} \bigl(\left\langle 1, 1 \middle\vert R^{3}\middle\vert 3,0 \right\rangle ^*\left\langle 0 ,1 \middle\vert R^{4}\middle\vert 4,0 \right\rangle +\left\langle 0, 1 \middle\vert R^{4}\middle\vert 4,0 \right\rangle ^*\left\langle 1 , 1 \middle\vert R^{3}\middle\vert 3,0 \right\rangle \bigr) \\
   &-9 \sqrt{21} \bigl(\left\langle 1, 1 \middle\vert R^{4}\middle\vert 4,0 \right\rangle ^*\left\langle 0 ,1 \middle\vert R^{3}\middle\vert 3,0 \right\rangle +\left\langle 0, 1 \middle\vert R^{3}\middle\vert 3,0 \right\rangle ^*\left\langle 1 , 1 \middle\vert R^{4}\middle\vert 4,0 \right\rangle \bigr)
   \Bigr).
 \end{split}
\end{equation}

\bibliography{TViolation,ParityViolation}
\end{document}